\documentclass[a4paper,11pt,pstricks]{article}
\usepackage{subfigure}
\pdfoutput=1 

\usepackage{jinstpub} 
\usepackage{pict2e}
\usepackage{calc}
\usepackage{fp} 
\usepackage{xcolor}
\usepackage{cleveref}
\usepackage{gensymb} 
\usepackage{rotating}

\usepackage{tikz}
\usepackage{varwidth} 
\usetikzlibrary{calc,trees,positioning,arrows,chains,shapes.geometric,%
    decorations.pathreplacing,decorations.pathmorphing,shapes,%
    matrix,shapes.symbols,decorations.text,shadows,shapes,shapes.geometric}

     \tikzset{
    device/.style={
    rectangle,
    rounded corners, 
    draw=black, thick,
    minimum height=0.5cm, 
    minimum width=2cm, 
    align=center,
    anchor=west,
    fill=white
    }
    }
\usepackage{import}
\usepackage{lineno}
\title{\boldmath The NA62 GigaTracKer: a low mass high intensity beam 4D tracker with \vu{65}{ps} time resolution on tracks}

\newcounter{savefootnote}
\newcounter{symfootnote}
\newcommand{\symfootnote}[1]{%
   \setcounter{savefootnote}{\value{footnote}}%
   \setcounter{footnote}{\value{symfootnote}}%
   \ifnum\value{footnote}>8\setcounter{footnote}{0}\fi%
   \let\oldthefootnote=\thefootnote%
   \renewcommand{\thefootnote}{\fnsymbol{footnote}}%
   \footnote{#1}%
   \let\thefootnote=\oldthefootnote%
   \setcounter{symfootnote}{\value{footnote}}%
   \setcounter{footnote}{\value{savefootnote}}%
}

\author[a]{G. Aglieri Rinella}
\author[a]{D. Alvarez Feito}
\author[e]{R. Arcidiacono}
\author[e]{C. Biino}
\author[a]{S. Bonacini}
\author[a]{A. Ceccucci}
\author[c]{S. Chiozzi}
\author[b]{E. Cortina Gil}
\author[c]{A. Cotta Ramusino}
\author[a]{H. Danielsson}
\author[a]{J. Degrange}
\author[a,b,c,d]{M. Fiorini}
\author[a]{L. Federici}
\author[c,d,a]{E. Gamberini}
\author[c]{A. Gianoli}
\author[a]{J. Kaplon}
\author[b]{A. Kleimenova}
\author[a]{A. Kluge}
\author[c,d]{R. Malaguti}
\author[a]{A. Mapelli}
\author[e]{F. Marchetto}
\author[a,b]{E. Mart\'in Albarr\'an} 
\author[e]{E. Migliore}
\author[b]{E. Minucci}
\author[a]{M. Morel}
\author[a]{J. No\"el}
\author[a]{M. Noy}
\author[b]{ G. N\"uessle} 
\author[a]{L. Perktold}
\author[a,b,1] {M. Perrin-Terrin \symfootnote{Corresponding author.} }
\note{present address: Aix Marseille Univ, CNRS/IN2P3, CPPM, Marseille, France}
\author[a]{P. Petagna}
\author[c,d]{F. Petrucci}
\author[a]{K. Poltorak}
\author[a]{G. Romagnoli}
\author[a,2]{G. Ruggiero}
\note{present address: Lancaster University, UK}
\author[b,3]{B. Velghe}
\note{present address: TRIUMF, Vancouver, British Columbia, Canada}
\author[d]{H. Wahl}

\affiliation[a]{CERN, Switzerland}
\affiliation[b]{UCL Louvain, Belgium}
\affiliation[c]{INFN Sezione di Ferrara,Italy}
\affiliation[d]{Universit\`a di Ferrara, Italy}
\affiliation[e]{INFN sezione di Torino, Italy}

\emailAdd{mathieu.perrin-terrin@cern.ch}

\newcommand{\fig}[1]{Figure~\ref{#1}}
\newcommand{\figs}[1]{\Cref{#1}}
\newcommand{\tab}[1]{Table~\ref{#1}}

\newcommand{\vu}[2]{\ensuremath{#1\,\unit{#2}}}
\newcommand{\veu}[3]{\ensuremath{(#1\pm#2)\,\unit{#3}}}

\newcommand{\unit}[1]{\ensuremath{{\rm{#1}}}}

\newcommand{\GeVc}{\unit{GeV/c}}

\newcommand{\MmGeV}{\unit{GeV^2/c^4}}

\newcommand{\nm}{\unit{nm}}
\newcommand{\sqmm}{\unit{mm^2}}

\newcommand{\mum}{\unit{\upmu m}}
\newcommand{\sqmum}{\unit{\upmu m^2}}
\newcommand{\murad}{\unit{\upmu rad}}
\newcommand{\mrad}{\unit{mrad}}

\newcommand{\mus}{\unit{\upmu s}}
\newcommand{\ns}{\unit{ns}}

\newcommand{\BR}{\ensuremath{\mathcal{B}}}
\newcommand{\BRof}[1]{\ensuremath{\mathcal{B}(#1)}}
\newcommand{\decay}[2]{\ensuremath{ #1 \to #2}}
\newcommand{\decayc}[2]{\ensuremath{ #1 {\to} #2}}
\newcommand{\particle}[1]{\ensuremath{\sf{#1}}}

\newcommand{\Kp}{{\particle{K^+}}}

\newcommand{\Pip}{{\particle{\pi^+}}}
\newcommand{\Piz}{{\particle{\pi^0}}}
\newcommand{\Pim}{{\particle{\pi^-}}}

\newcommand{\Kpinn}{\decay{\Kp}{\Pip \nu \overline{\nu}}}

\newcommand{\Kpipizero}{\decay{\Kp}{\Pip \pi^0 }}

\newcommand{\Kpipipi}{\decay{\Kp}{\Pip \Pip \Pim }}

\abstract{
The GigaTracKer (GTK) is the beam spectrometer of the CERN NA62 experiment. The detector features challenging design specifications, in particular a peak particle flux reaching up to \vu{2.0}{MHz/mm^2}, a single hit time resolution smaller than \vu{200}{ps} and, a material budget of \vu{0.5\%}{X_0} per tracking plane. To fulfil these specifications, novel technologies were especially employed in the domain of silicon hybrid time-stamping pixel technology and micro-channel cooling. This article describes the detector design and reports on the achieved performance.

}


\arxivnumber{1904.12837} 

\begin{document}
 \maketitle
  \flushbottom

  \section{Introduction} \label{sec:intro}
  
The GigaTracKer (GTK) is the silicon time-resolving hybrid pixel beam spectrometer of the NA62 experiment. The NA62 experiment~\cite{hep-ph_NA62_2017c} is installed in one of the north area extraction lines, called P42, of the CERN Super Proton Synchrotron (SPS). The experiment main goal is to measure the branching fraction (\BR) of the ultra rare \Kpinn\ decay, $$\BRof\Kpinn = (8.4\pm 1) \times 10^{-11},$$ as predicted by the Standard Model (SM)~\cite{hep-th_BurasEtAl_2015}, with a 10\% precision~\cite{hep-ph_NA62_2018a}. This observable provides remarkably stringent tests for the SM and the physics models beyond it~\cite{hep-th_BurasEtAl_2014}.

\begin{figure}[h]\begin{center}
  \includegraphics[width = \textwidth]{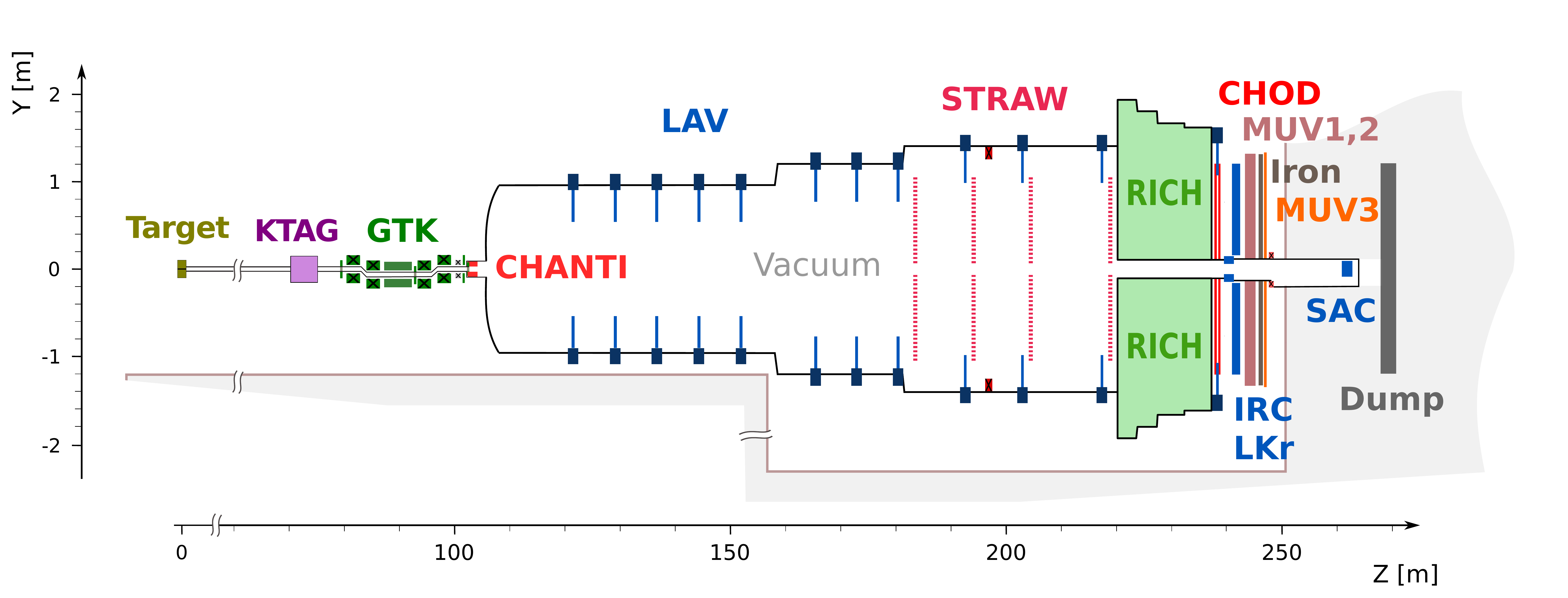}
  \caption{Schematic longitudinal view of the NA62 experiment.
    The SPS proton beam impinges on a target producing a secondary beam of \Pip , proton and \Kp. \Kp\ are identified by a differential Cerenkov detector (KTAG). Beam particles momentum and direction are measured by the GTK. A plastic scintillator based guard ring (CHANTI) is used to veto inelastic collisions occurring in the GTK last station. A fast plastic scintillator based detector (CHOD) is used to trigger the experiment readout when a charged decay product of a beam particle is detected. The decay products are reconstructed with a straw tube based magnetic spectrometer (STRAW), a set of electromagnetic calorimeters (LAV, LKr, IRC and SAC), a Cerenkov detector (RICH), a set of hadronic calorimeters (MUV1 and MUV2) and a plastic scintillator based muon veto (MUV3). See~\cite{hep-ph_NA62_2017c} for more details.}
  \label{fig:NA62Schematics}
\end{center}\end{figure}

Measuring the branching fraction of an ultra rare decay in such a final state is experimentally challenging. In the previous experiments dedicated to this observable, E787 and E949~\cite{hep-ph_E949_2009}, a kaon beam was stopped in a target surrounded by detectors measuring the properties of the particles produced by the kaons decaying at rest. At NA62, a decay in flight technique is employed. A beam of \veu{75}{1}{\GeVc} particles, composed of positively charged \Kp (6\%), \Pip (70\%) and \particle{p} (23\%) with a total nominal rate of \vu{750}{MHz}, is transported through a \vu{270}{m} long evacuated tank in which the particles may decay. The tank is instrumented to measure the properties of the beam particles and the decay products, as shown in \fig{fig:NA62Schematics}. The momentum, direction, spacial and time coordinates of the beam particles are measured using the GTK. To allow for such a measurement, the beam is focussed and collimated to a transverse size of around \vu{6\times3}{cm^2} and a maximum particle flux of \vu{2.0}{MHz/mm^2}.

\begin{figure}[!b]\begin{center}
  \includegraphics[width = \textwidth]{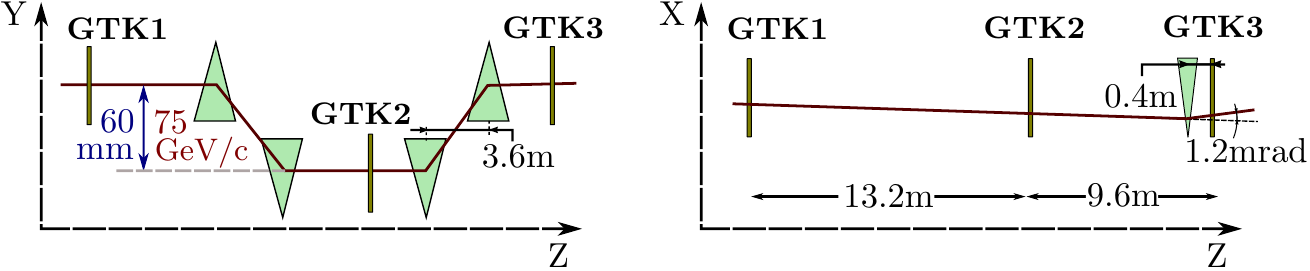}
  \caption{Schematic view of the GTK. The green triangles represent bending magnets. The beam is deflected by \vu{1.2}{\mrad} in the XZ plane to compensate the effect of the STRAW spectrometer dipole magnet.}
  \label{fig:GTKSchematics}
\end{center}\end{figure}

The GTK is composed of three stations, each made of 18000 time resolving pixels of \vu{300 \times 300}{\sqmum}, covering a total area of \vu{60.8\times27}{\sqmm}. The stations are installed in the beam pipe in vacuum around two pairs of bending magnets as shown in \fig{fig:GTKSchematics}.
The magnet pairs displace the beam particles vertically such that they describe, between the two magnet pairs, a trajectory parallel to the undeviated one, and, that they emerge, after the last magnet, as undeviated with respect to the original trajectory. A fifth magnet deflects the beam by \vu{1.2}{\mrad} in the XZ plane to compensate the horizontal displacement due to the STRAW spectrometer dipole magnet.
The positions recorded in the first and third GTK stations allow the measurement of the particle direction. This information, together with the position measured in the second GTK station, allow the evaluation of the vertical displacement undergone between the magnet pairs. This displacement depends on the momentum rigidity and amounts to \vu{60}{mm} for \vu{75}{GeV/c} particles.

\begin{figure}[b]\begin{center}
\resizebox{.7\textwidth}{!}{
  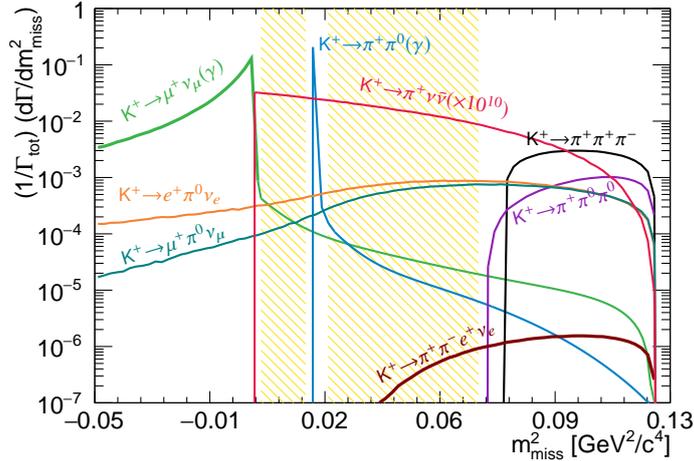}
  \caption{Squared missing mass distribution of the main kaon decay modes and \Kpinn\ enhanced by 10 orders of magnitude. The two signal regions are represented by the dashed area.}
  \label{fig:M2miss}
\end{center}\end{figure}

The beam particles direction and momentum are used to reconstruct the signal candidates squared missing mass, 
\begin{equation} \label{eq:mm2}
m_m^2 = |\bold{p}_\Kp-\bold{p}_\Pip|^2,
\end{equation}
where $\bold{p}_{\Kp (\Pip)}$ denotes the 4-momentum of the parent (decay) particles assumed to be a \Kp\ (\Pip). The \Kpinn\ signal is searched in the two squared missing mass regions on either side of the peak of the \Kpipizero\ decay, displayed as dashed areas in \fig{fig:M2miss}. With a branching fraction of $20.67 \pm 0.08 \%$~\cite{hep-ph_ParticleDataGroup_2018}, this decay is 9 to 10 orders of magnitude more abundant than the \Kpinn. Measuring the signal branching ratio with a 10\% precision requires therefore a \Kpipizero\ suppression of 11 orders of magnitude. This suppression can be achieved with an efficient \Piz\ rejection and an event selection based on the decay squared missing mass. The former provides a background rejection of 7 orders of magnitude~\cite{hep-ph_NA62_2017c}, requiring the latter to be of 4 orders of magnitude. To obtain this rejection, based on simulations, the resolution on the \Kpipizero\ squared missing mass is required to be smaller then \vu{0.001}{\MmGeV}. This requirement corresponds to an angular and momentum resolution on the beam particle at the exit of the GTK of $\theta_{x,y} =p_{x,y}/p_z = \vu{16}{\murad}$ and $\delta p / p = 0.2\%$.

This performance constrains the maximum amount of material crossed by the beam in the GTK as shown in \fig{fig:GTKExpectedKineReso}. Beside this loose constraint, the GTK material budget is, above all, critical to control the number of inelastically scattered beam particles. Indeed, while the four magnets are sweeping away fragments produced by particles inelastically scattered in the first two GTK stations, those produced in the third station may enter the acceptance of the other detectors and generate candidates mimicking the signal. Extrapolating from existing technologies~\cite{hep-tec_ALICESPD_2005}, an approximative limit on the amount of material crossed by the beam at each station of 0.5\% of a radiation length (\unit{X_0}) or 0.08\% of a nuclear interaction length ($\lambda_i$) was considered achievable.

The pixel time resolution is essential to reconstruct the particle tracks inside the GTK. 
Indeed for each kaon decay considered as valuable by the trigger system, a GTK readout spanning over \vu{75}{ns} is recorded. Considering the instantaneous nominal beam particle rate of about 0.75 particles/ns, a hit list from more than 50 particles is expected for each record.
As the three-station configuration offers almost no geometrical redundancy, track pattern recognition relies mostly on hits time coordinate comparison. With \vu{200}{ps} time resolution per plane, hits from two consecutive particles are separated by a duration larger than five times the hit resolution. 
As a result, the resolution on the beam particle arrival time is better than \vu{150}{ps} which allows unambiguous association of the GTK track with the activities measured in other subdetectors in particular the RICH detector which feature a \vu{90}{ps} time resolution.

Nonetheless, wrong associations are still possible if some particles are not reconstructed due to detector inefficiency. If such mis-associations occur for a \Kpipizero\ decay, the squared missing mass obtained will be incorrect and ineffective to reject the event. 
The probability for these events to occur is the product of the GTK inefficiency by the probability to have a beam pion or proton forming a good vertex (i.e. with a tracks distance of closest approach and a time difference both within 3 times the detector resolution) with the \Kp\ decay product. The later being 0.3\%, the GTK inefficiency has therefore to be below 3\% (corresponding to a station efficiency above 99\%) to reach a mis-association probability below $10^{-4}$ and thus compensating for the inefficiency of the squared missing rejection of \Kpipizero\ (nominally 4 orders of magnitude).

Finally, at a beam particle rate of \vu{750}{MHz}, for an average year of 200 days of operation with a duty cycle of the accelerator of 25\%, the beam deposits in the beam center a peak (average) dose of \vu{15.5~(4.5)}{MRad} corresponding to a fluence of \vu{4.5~(1.0)\times10^{14}}{1MeV n_{eq}/cm^2}. 
Despite the irradiation (larger than initially expected~\cite{hep-ph_AnelliEtAl_2007} due to revisions of the experimental conditions) the performance previously mentioned and reported in \tab{tab:requirements} has to be maintained.

\begin{table}[h!]
\caption{Specifications for the GigaTracKer at nominal beam intensity. At 65\% nominal beam intensity, corresponding to the maximum operating intensity in 2016 and 2017, the specifications for the rate, flux and fluence are 65\% of those reported in the table. }
\label{tab:requirements}
 \begin{center}
\begin{tabular}{|l|r|}\hline
  Beam Particle Rate & \vu{750}{MHz}\\
  Peak Beam Particle Flux & \vu{2.0}{MHz/mm^2}\\\hline
	     Average Fluence & \vu{1.0 \times 10^{14}}{1\,MeV~n_{eq} /cm^{2}/200~ days}\\
     	     Peak Fluence & \vu{4.5 \times 10^{14}}{1\,MeV~n_{eq} /cm^{2}/200~ days}\\\hline
		Average Integrated Dose & \vu{3.5}{MRad / 200~days}\\
		Peak Integrated Dose & \vu{15.5}{MRad / 200~days}\\\hline
	    Station Efficiency & 99\%\\\hline
	    Momentum Resolution & 0.2\%\\\hline
	    Angular Resolution & \vu{16}{\murad}\\\hline
	    Pixel Time Resolution & \vu{200}{ps} RMS\\\hline
	    Station Material Budget & \vu{0.5\%}{X_0}\\\hline
	  \end{tabular}	  
\end{center}
\end{table}

  \section{Detector Design}\label{sec:design}

\begin{figure}[h]
	\subfigure[]{
		\includegraphics[height=4.2cm]{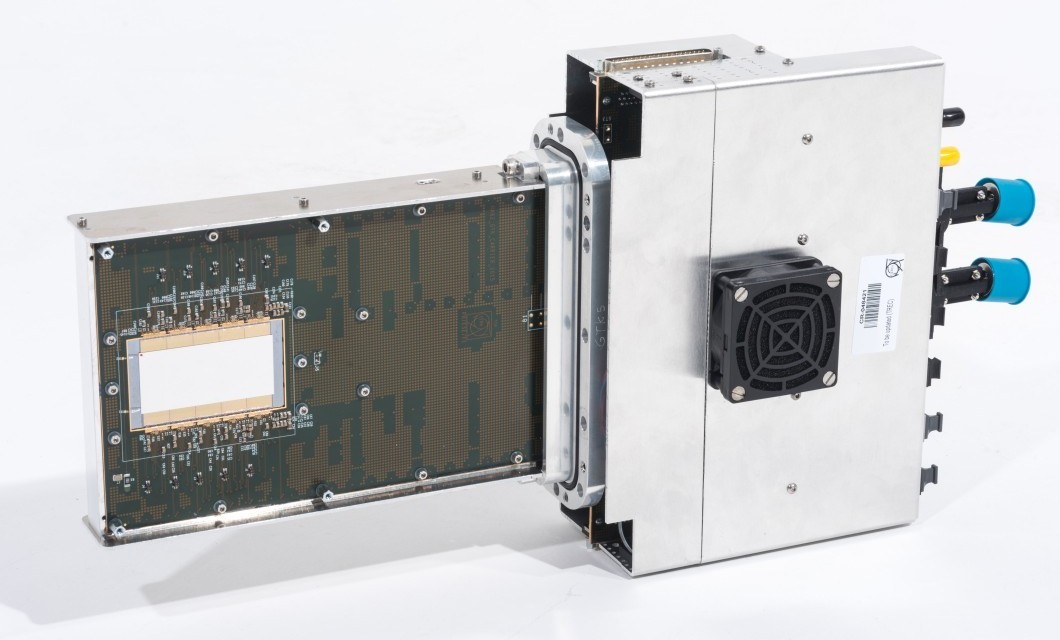}
		\label{fig:GTKPhotosSensor}
	}
	\subfigure[]{
		\includegraphics[height=4.2cm]{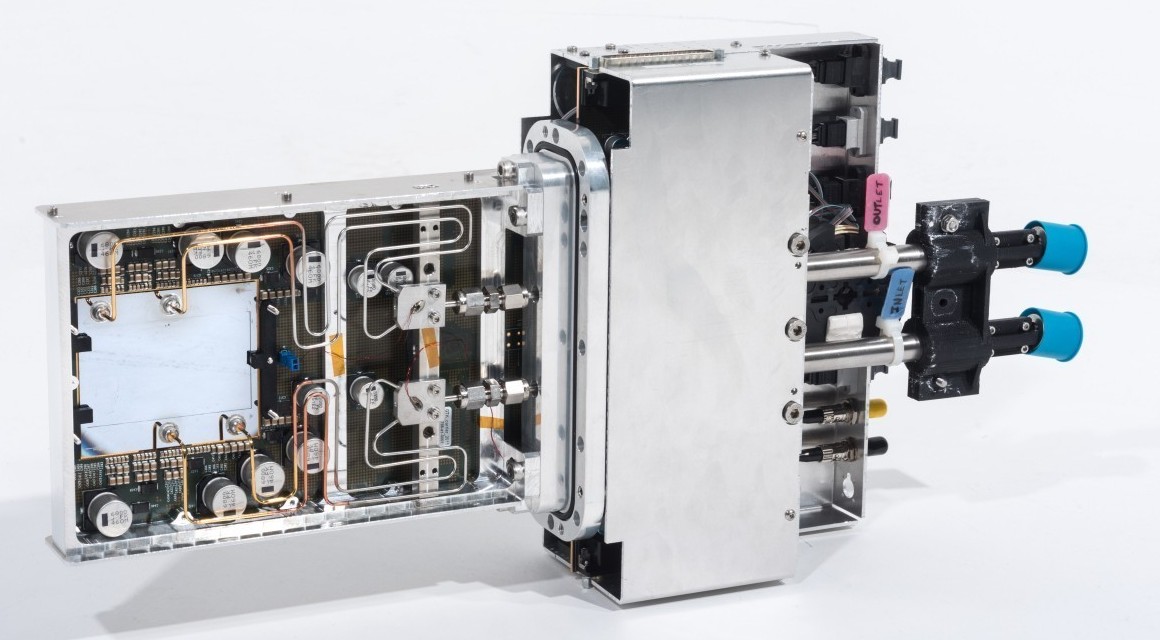}
		\label{fig:GTKPhotosCooling}
	}
	\caption{Photograhs of a GTK station viewed from the sensor side \protect\subref{fig:GTKPhotosSensor} and from the cooling plate side \protect\subref{fig:GTKPhotosCooling}.}
	\label{fig:GTKPhotos}
\end{figure}

The GTK was designed to fulfil the specifications listed in \tab{tab:requirements}. 
The irradiation level has strong implications on the overall design of the detector. Indeed, the GTK had to be made such that any tracking plane could be promptly replaced once the irradiation had induced significant performance degradation. This consideration imposed a tracking station concept that would be compact, standard and self-contained.
Photographs of the station are shown in \fig{fig:GTKPhotos}.

\begin{figure}[h]
	\begin{center}
		\includegraphics[width = \textwidth]{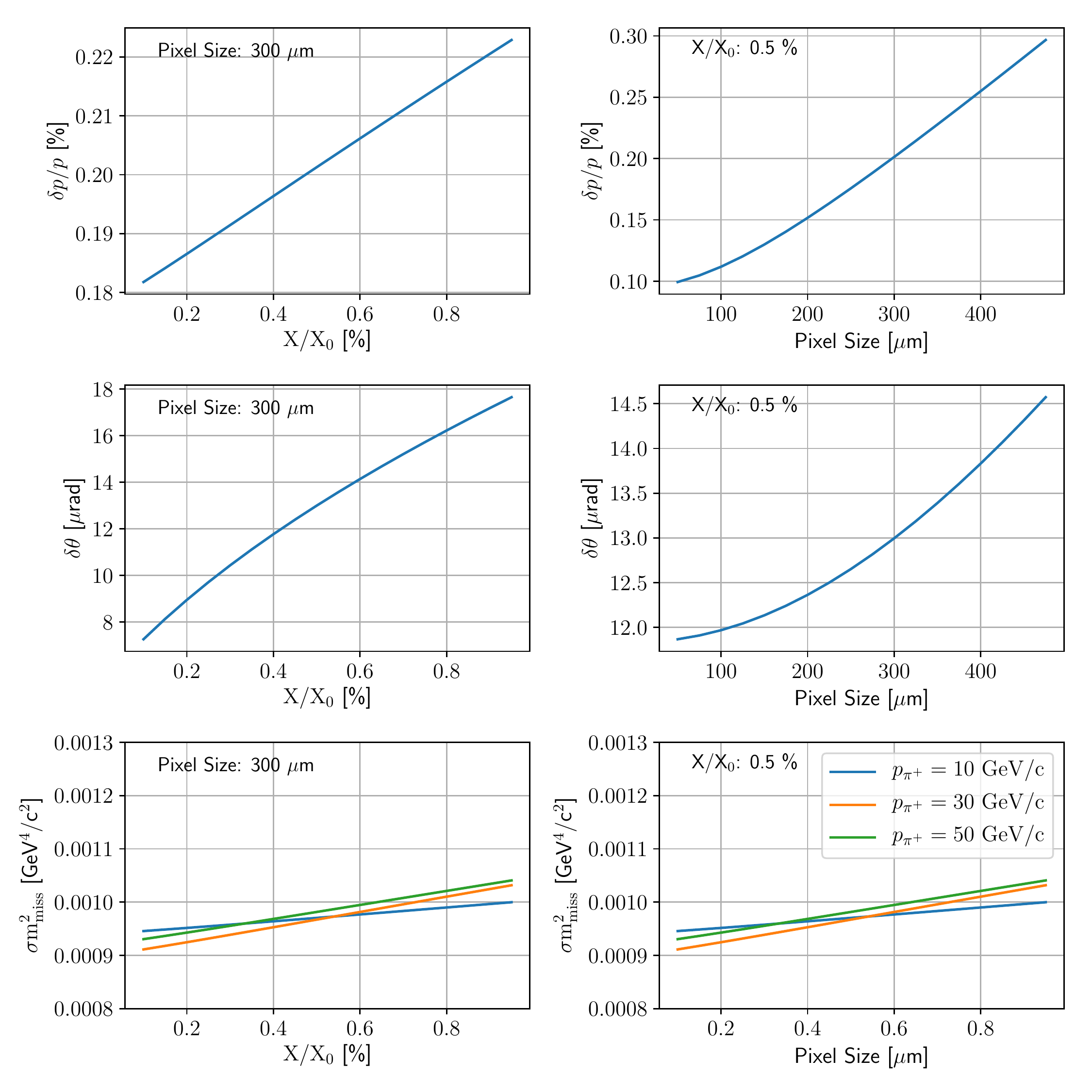}
		\caption{Expected momentum, angular and \Kpipizero\ squared missing mass resolution as a function of the material budget and pixel size. The \Kpipizero\ squared missing mass resolution is given for different \Pip\ momenta.}
		\label{fig:GTKExpectedKineReso}
	\end{center}
\end{figure}

The \vu{60.8\times27}{\sqmm} sensitive region of the detector is made of a  \vu{200}{\mum} thick silicon sensor read out by two rows of five custom made application specific integrated circuit (ASIC) called TDCPix, thinned to \vu{100}{\mum}. The choice of the chip pixel size results from a compromise between the number of channels to read out, the particle rate per channel and the angular and momentum resolution. Using \vu{300\times300}{\sqmum} pixels, 18000 channels are needed to cover the sensitive region and \vu{180\times 10^3}{particle/s} cross the most central pixels. The expected kinematics performance with such pixels are within the specifications and shown as a function of the pixel size in \fig{fig:GTKExpectedKineReso}.

The hybrid pixel detector (HPD) made by the sensor bonded to the 10 chips is glued on a silicon plate serving both as mechanical support and heat exchanger. The plate is etched to form channels with a cross section of \vu{200\times70}{\sqmum} in which a coolant (liquid \unit{C_6F_{14}}) is circulated to convect the heat produced by the ASICs. In order to minimise the amount of particles inelastically scattered in the detector, the cooling plate thickness was minimised. A thickness of \vu{210}{\mum} was considered technologically achievable. As such, the material budget of the assembly made by the HPD and the cooling plate is about \vu{0.5\%}{X_0}. \fig{fig:GTKExpectedKineReso} shows that such material budget allows the achievement of the required kinematics performance.

\begin{figure}
	\begin{center}
		\resizebox{.7\textwidth}{!}{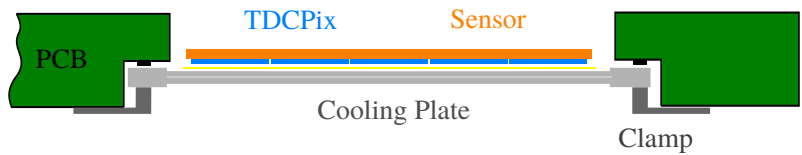} 
	\end{center}
	\caption{Schematics cross view of the assembly made of the HPD and cooling plate inserted in the PCB Carrier.}
	\label{fig:GTKSchematicsCrossView}
\end{figure}
			
The assembly made of the HPD and cooling plate is inserted in a countersink in a printed circuit board (Carrier PCB) to which the plate is clamped as shown in \fig{fig:GTKSchematicsCrossView}. The chips are electrically connected with aluminium wire bonds to the PCB to transfer power, clocks, configuration instructions and output data. The PCB is glued into a frame and a vacuum flange such that, once mounted in the beam pipe vessel, the board serves as vacuum feedthrough. In the region outside the vacuum vessel, the PCB is equipped with optical tranceivers communicating through fibers with readout boards 250 meters away in surface counting room. These boards provide clocks and configuration instructions and receive and process the data of all fired pixels. The readout boards are connected to the data acquisition (DAQ) system of the experiment. Upon request of this system, the boards retrieve the information of all the pixels fired around a given reference time and send them to a computer farm which assembles these data. These computers finally send the assembled data packets to the NA62 central computer farm which, in turn, assembles information from all detectors. The next sections describe in details each of the elements constituting the GTK.

\subsection{Sensor}
			
The relevant constraints for the design of the GTK sensor are, first, the intense radiation exposition with a fluence reaching \vu{2\times10^{14}}{1MeV n_{eq}/cm^2} in 100 days of run and second, the minimization of the station total material budget fixed to a maximum of \vu{0.5}{X_0}.
			
As far as the latter is concerned, a sensor thickness of \vu{200}{\mum} was found to be optimal. Such a thickness limits the sensor contribution to the material budget to 0.2\% \unit{X_0} and still provides a sufficiently large signal: 15000 electron-hole pairs corresponding to a most probable charge value of \vu{2.4}{fC} for a \vu{75}{\GeVc} particle.
			
The exposure to radiation has two main consequences: the increase of the sensor leakage current and the reduction of the sensor charge collection efficiency. Both effects can be mitigated by cooling the sensor. Moreover, the loss of charge collection efficiency can be recovered by increasing the bias voltage. At the sensor end-of-life, the bias voltage was foreseen to be raised up to \vu{750}{V}.
Finally, n-in-p type sensors are in principle more resistant to radiation as they can still operate under partial depletion. However, the use of these sensors was still uncertain at the time of the GTK design, hence both p-in-n and n-in-p sensor types were manufactured.

\begin{figure}[b]
	\begin{center}
		\includegraphics[width =.7\textwidth]{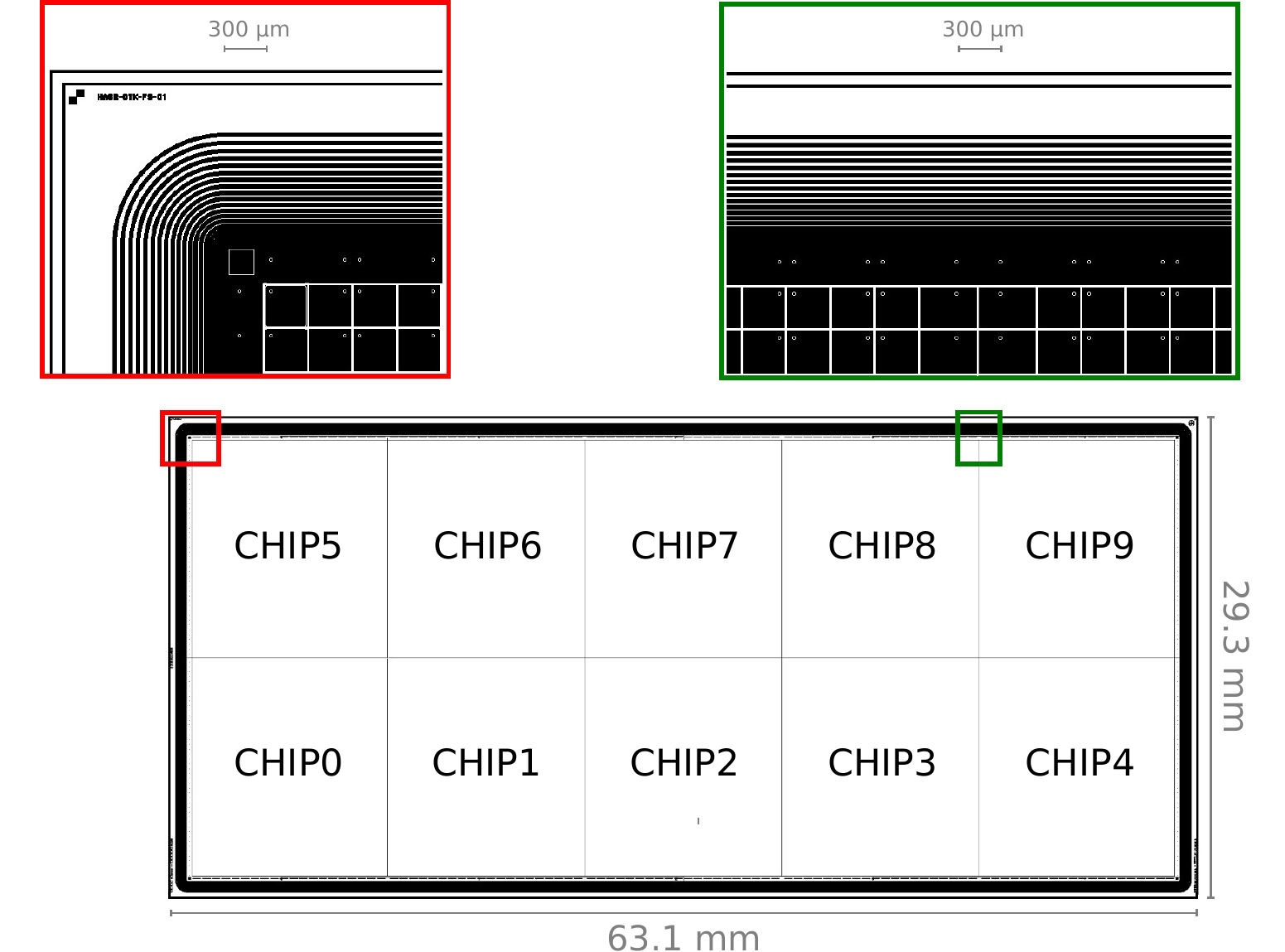}
		\caption{GTK sensor geometry: the top zoomed inlets show the guard rings and the pixel matrix at the corner (left, red) and at the transition between two TDCPix (right green) where enlarged pixels are placed.}
		\label{fig:sc-sensor}
	\end{center}
\end{figure}		
Prototype sensors made of \vu{3\times 3}{mm^2} n-in-p and p-in-n diodes were irradiated up to a fluence of \vu{2\times10^{14}}{1MeV n_{eq}/cm^2}, corresponding to 100 days of operation, and showed a satisfying performance. The sensor leakage currents and depletion voltages were measured before and after irradiation. The maximum full depletion voltage after 100 days of operation was measured to be around \vu{300}{V} while the sensors were stably operated at \vu{750}{V}. The leakage current measurements showed that with a sensor temperature of 5\textdegree\ or less, the sensors can safely be operated for 100 days.

The sensor geometry, shown in \fig{fig:sc-sensor}, is designed to be bump-bonded to two rows of five TDCPix and features extra horizontal spacing between the bonding pads at the chip borders thus forming enlarged pixels (\vu{400\times300}{\mum^2} instead of \vu{300\times300}{\mum^2}).  The bump-bonding employs SnAg solder bumps with a diameter of \vu{30}{\mum} and a height of \vu{25}{\mum}. A layer of \vu{3}{\mum} of benzocyclobutene was deposited on both the sensor and the chips to prevent discharges between these two elements.

\subsection{TDCPix}\label{ssec:tdcpix}
A dedicated readout chip, the TDCPix~\cite{hep-tec_AglieriRinellaEtAl_2013,hep-tec_NoyEtAl_2011,hep-tec_KlugeEtAl_2013,hep-tec_RinellaEtAl_2015}, was designed for the GTK specifications. The chip main functionalities are the sensor hit signal amplification, discrimination, digitisation, time-stamping and the transmission of the resulting digitised data off chip. Most of the specifications reported in \tab{tab:requirements} apply directly to the chip design. The particle rate translates into a maximum pixel hit rate of \vu{180}{kHz} and implies a careful handling of radiation induced single event upset (SEU).
				
\begin{figure}
	\subfigure[]{ 
		\resizebox{!}{8cm}{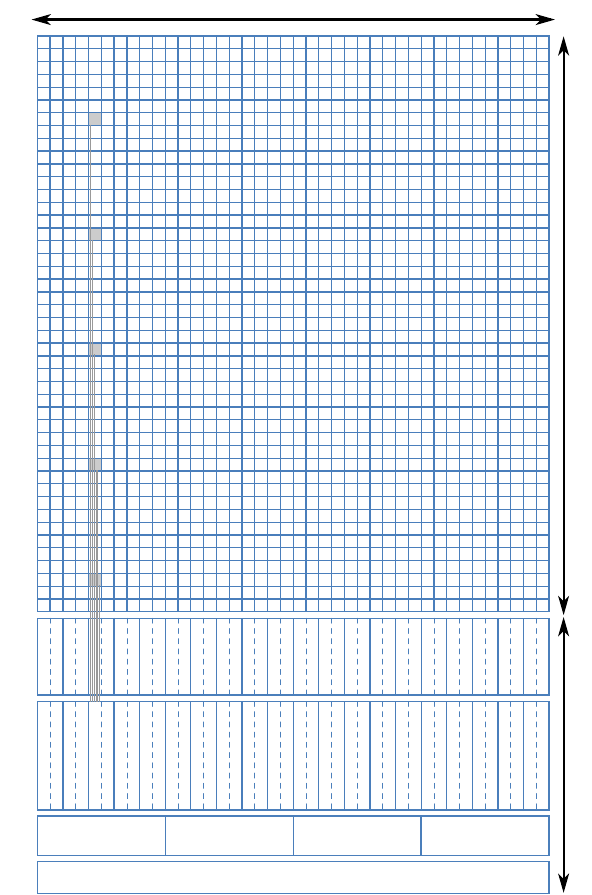}
		\label{fig:sc-chip-a}
	}
	\subfigure[]{
		\resizebox{!}{8cm}{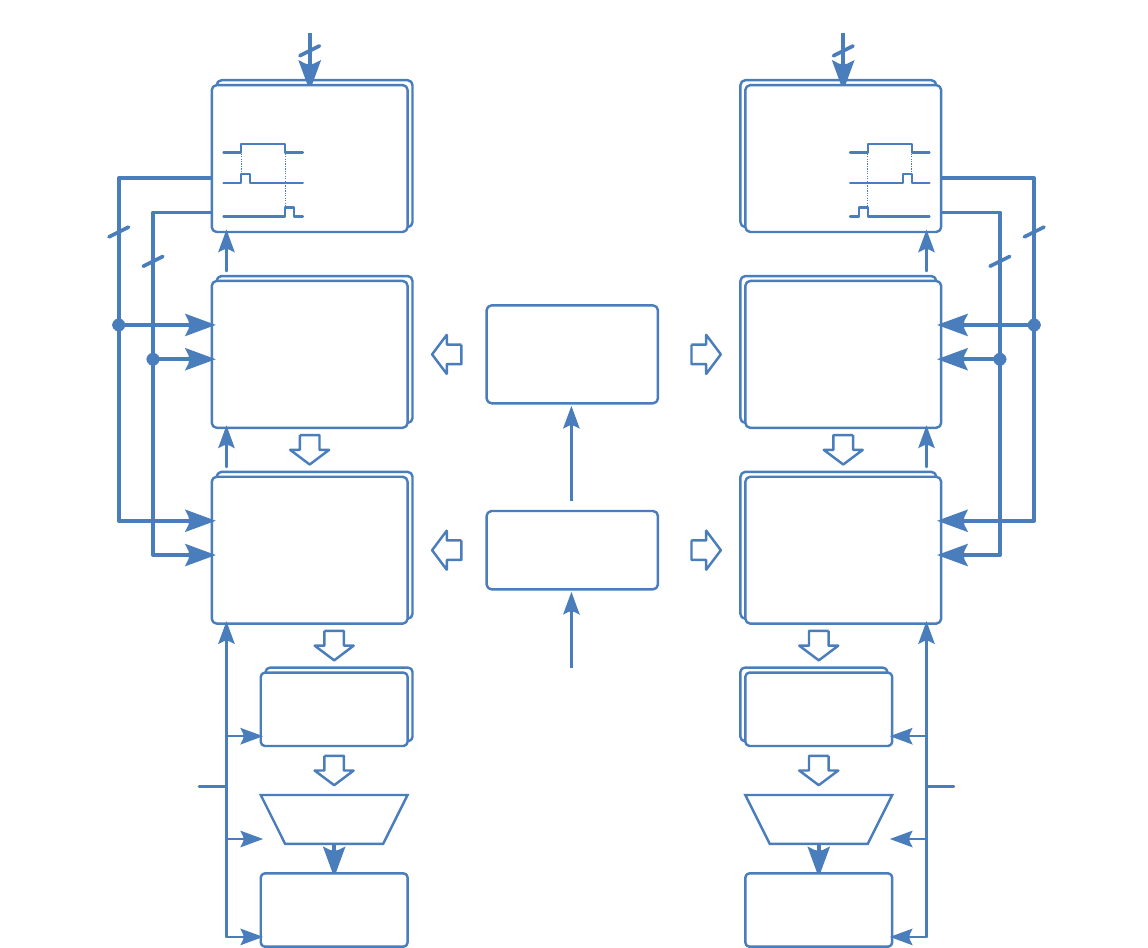}
		\label{fig:sc-chip-b}
	}
	\caption{\protect\subref{fig:sc-chip-a} The TDCPix architecture is separated in two parts, the pixel matrix and the end-of-column (EoC) regions. The output of five geometrically separated pixels are multiplexed in one time-stamping unit located in the EoC. The outputs of the  time-stamping units are sent out by four serialisers. \protect\subref{fig:sc-chip-b} Each time-stamping unit is fed by one clock shared amongst two pixel columns. This clock is used to increment a counter and to feed a delay-locked loop (DLL) dividing the clock period by 32. Hits leading and trailing edges are time-stamped by latching the counter and the DLL states in two registers whose content is encoded and multiplexed to be sent to the serialisers.}
	\label{fig:sc-chip}
\end{figure}

The TDCPix was designed in a commercial \vu{130}{\nm} technology. The chip is organised in two main parts: first, the 40 $\times$ 45 pixel matrix functioning asynchronously where the sensor hit signals are digitised and second, the end-of-column (EoC) area where those digitised hits are time-stamped and serialised. As such, the analog logic is separated from the digital one which reduces digital switching noise in the amplifying and digitisation circuits. A schematics of the chip is shown in \fig{fig:sc-chip-a}.

The expected electron collection time for a sensor bias voltage between \vu{100}{V} and \vu{750}{V} ranges from \vu{4}{ns} to \vu{2}{ns}. At the pixel, the hit induced current on the electrode is shaped by an amplifier with a \vu{5}{ns} peaking time and then followed by a Time-over-Threshold (ToT) edge discriminator whose threshold can be trimmed. Despite the variation of the electron collection time, the convoluted response of the amplifier and the input signal has a stable rise time (20-80\%) ranging from \vu{2.6}{ns} to \vu{2.1}{ns}.
				
\begin{figure}[b]
	\begin{center}
		\includegraphics[width=0.6\textwidth]{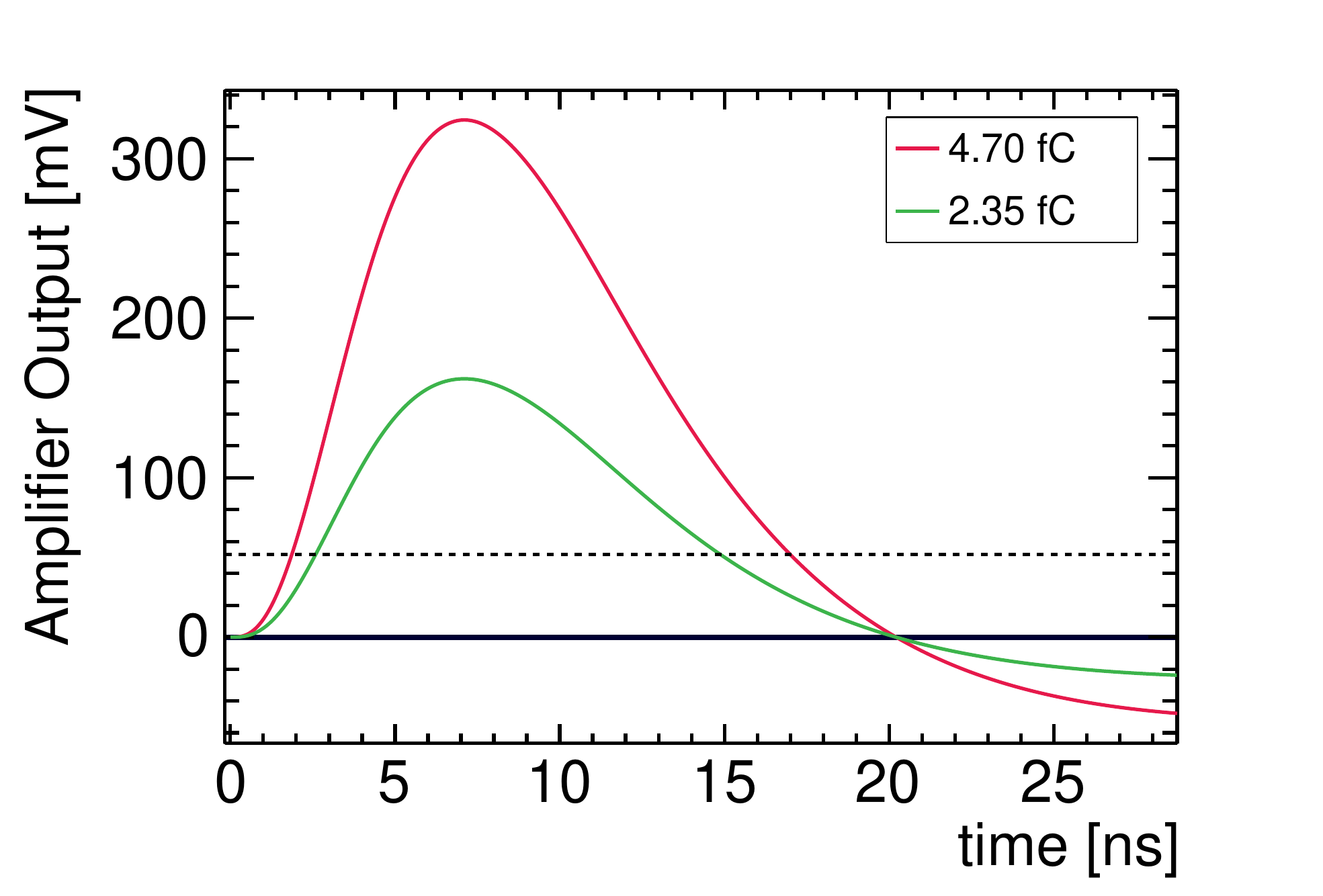}
		\caption{ Pixel amplifier output for \vu{100}{V} sensor bias, simulated using the WeightField2 software~\cite{hep-tec_CennaEtAl_2015}. 
		For a fixed hit arrival time, signals with larger amplitudes (red line) cross the discriminator threshold (dotted black line) earlier than smaller signals (green), inducing  a time-walk.}
		\label{fig:sc-tot}
	\end{center}
\end{figure}

At the discriminator output, the digital signal associated to the hit has a length corresponding to the ToT duration. For a fixed hit arrival time, signals with larger amplitudes cross the discriminator threshold earlier than smaller signals, inducing  a time-walk up to \vu{2.2}{\ns}, as shown in \fig{fig:sc-tot}. 
Being the ToT value directly related to the signal amplitude, it can thus be used to correct for the time-walk.
The digital hit signals are shipped on individual differential lines to the EoC using the pre-emphasis technique. Furthermore, groups of five pixels are multiplexed into nine time-stamping units as shown in \fig{fig:sc-chip-b}. These units consist of two time-to-digital converters (TDC), registering the signal leading and trailing edges. Each TDC is composed of a fine and a coarse time counter of \vu{100}{ps} and \vu{3.125}{ns} precision, respectively, with a range of up to \vu{6.4}{\mus}\footnote{The counters are incremented by clocks derived from a \vu{320}{MHz} clock. For the fine time counter, the clock is divided by 32 corresponding a \vu{97.65625}{ps} period.}. The hit time-stamp and pixel address are encoded into a 48 bit word stored in a buffer before being shipped out by four \vu{3.2}{Gb/s} serialisers each serving 10 pixel columns. To further extend the time-stamp range to 1718 seconds, the serialiser intercalates every \vu{6.4}{\mus} a frame word containing an incrementing 27 bit counter. Hit collisions in the multiplexer are kept below the percent level by grouping non-adjacent pixels. In case of collision, the time-stamping unit inputs are locked by the first incoming hit until it is processed. The concurrent hits are not time-stamped but the hits pixel addresses are encoded in the word of the first incoming hit. The configuration registers are triplicated to mitigate SEU effects.

The chip power dissipation is \vu{4.1}{W}  and varies across the chip, with approximately \vu{4.8}{W/cm^2} in the EoC and \vu{0.32}{W/cm^2} in the pixel matrix\footnote{The chip operating voltages are \vu{1.35}{V} for the EoC and \vu{1.3}{V} for the pixel matrix.}. As the detector is in vacuum, active cooling is therefore required.
					
\subsection{Micro-channel Cooling Plate}
					
The cooling system was designed to cool the front-end electronics and the sensor to less than \vu{5}{\degree C} with a maximum temperature difference of less than \vu{10}{ \degree C}. In addition, the amount of material in the beam acceptance has to be as small as possible. To fulfil these requirements a micro-channel cooling device was developed~\cite{hep-tec_RomagnoliEtAl_2015}. The implementation of such a device is the world's first for high energy physics. Using this technology, the cooler -- a silicon plate -- also serves as mechanical interface between the HPD and the carrier board. This additional functionality adds a constraint on the plate planarity to be better than \vu{30}{\mum}. As the cooler is entirely made of silicon, the coefficient of thermal expansion of the plate is naturally matched to the HPD one. 
					
\begin{figure}[bt]
	\setbox1=\hbox{\includegraphics[width = 0.3\textwidth]{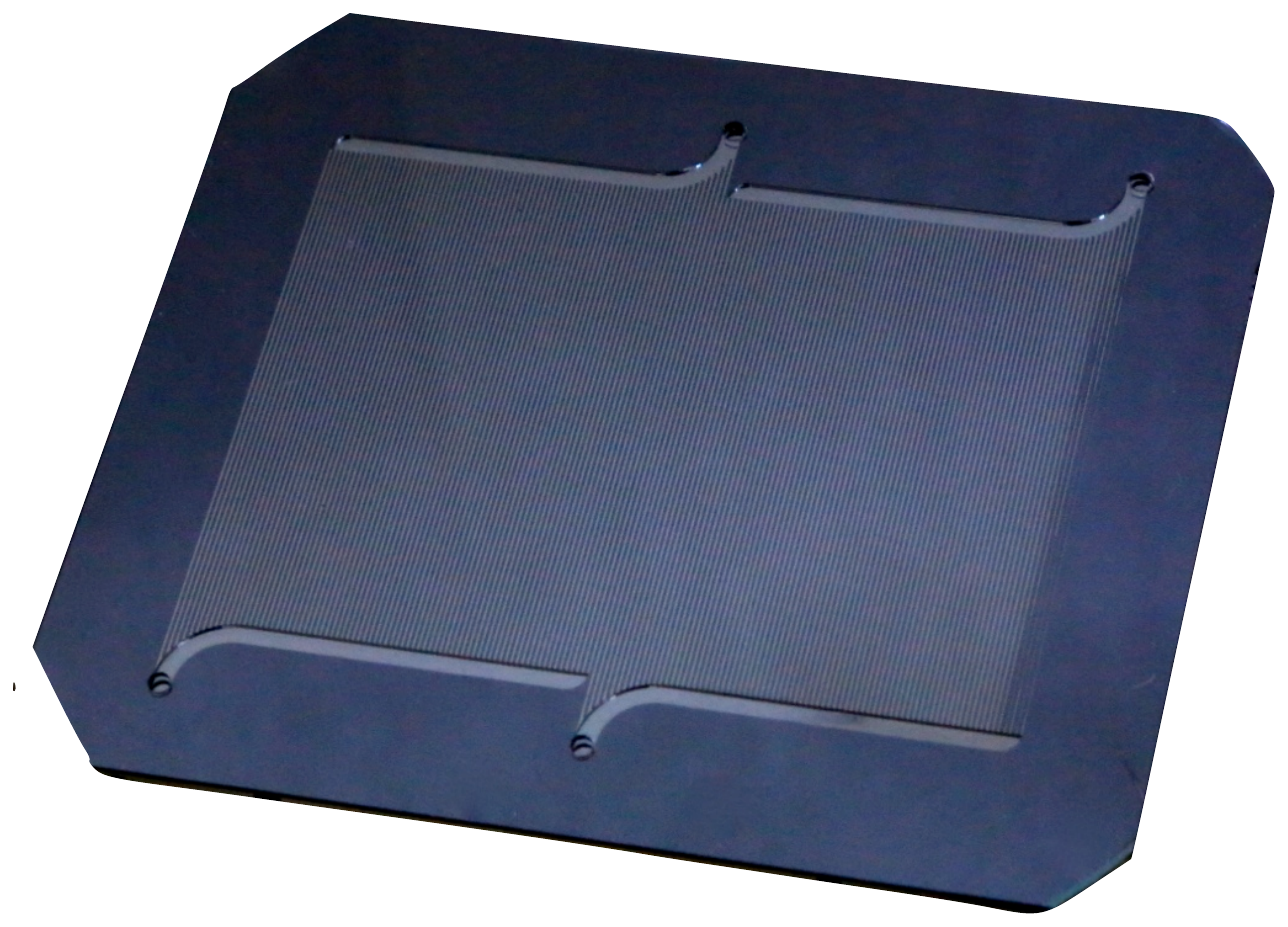}}
	\setbox2=\hbox{\includegraphics[width = 0.55\textwidth]{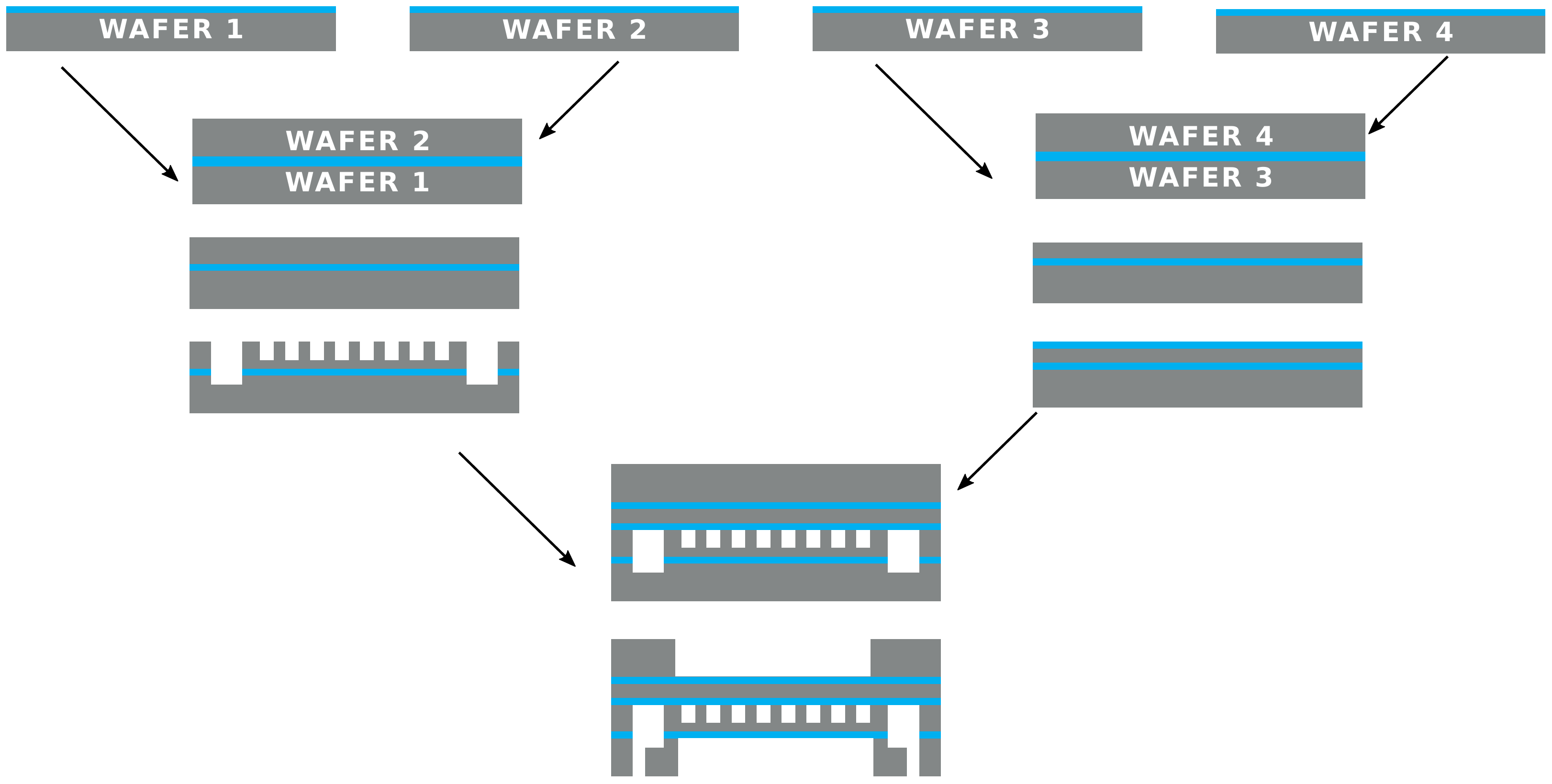}}
	\setbox3=\hbox{\includegraphics[width = 0.3\textwidth]{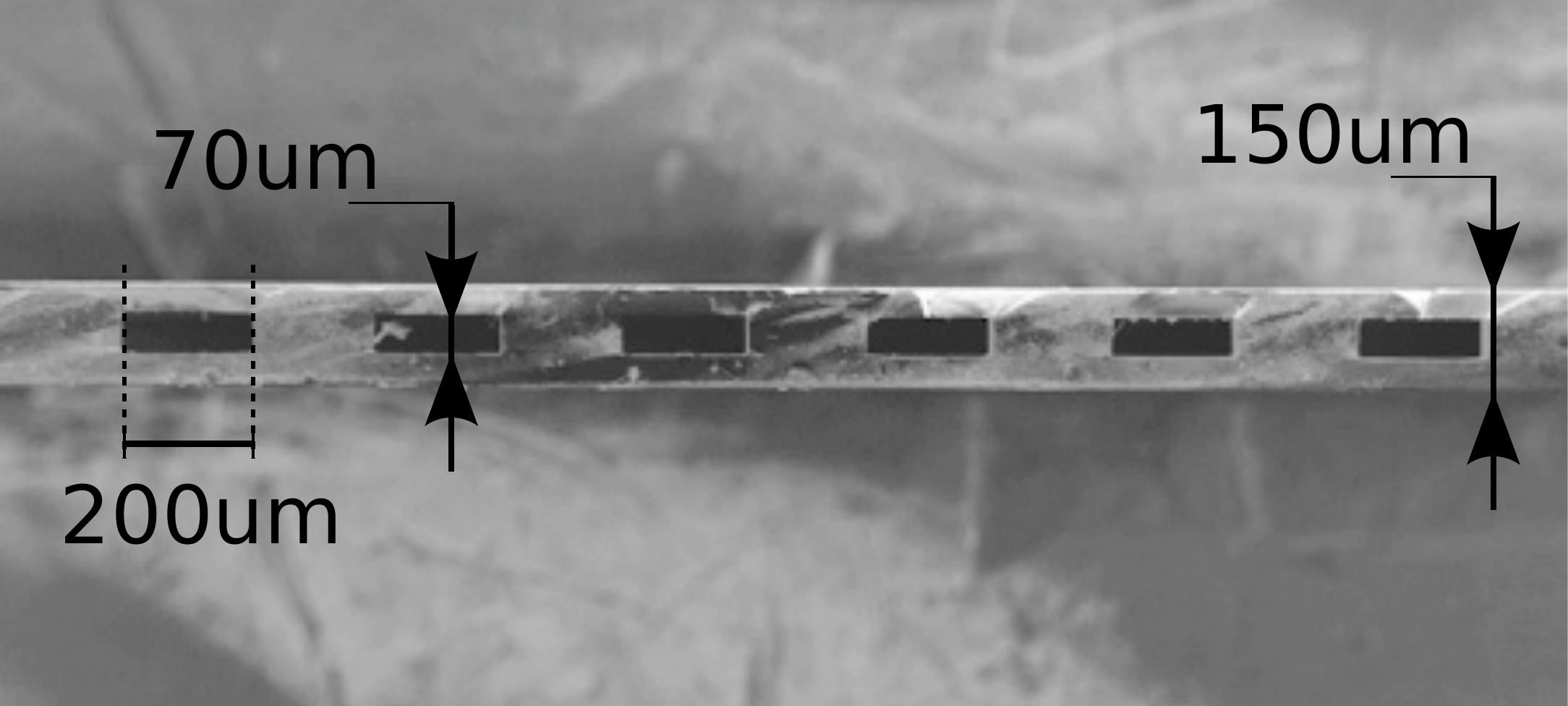}}
	\setbox4=\hbox{\includegraphics[width = 0.4\textwidth]{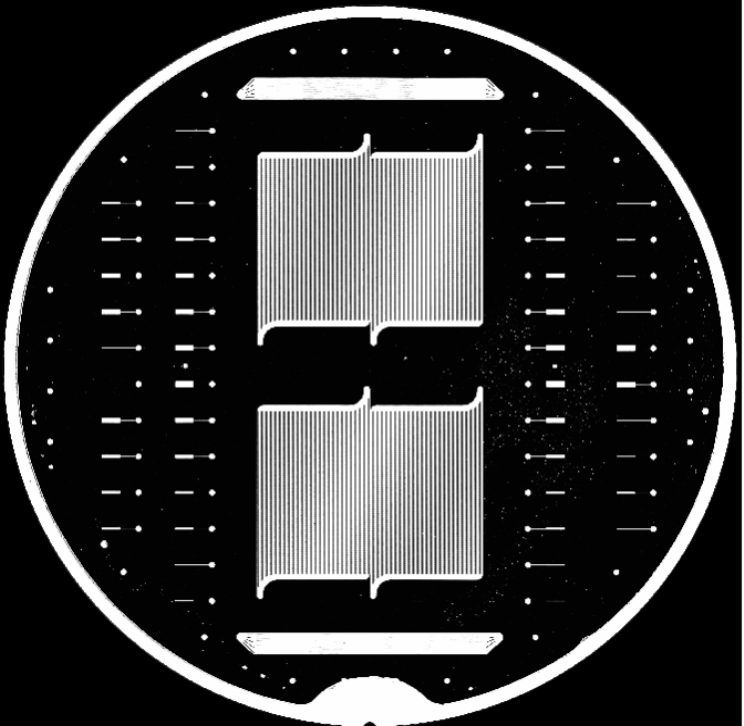}}
							
	\begin{center}
		\subfigure[]{
			\raisebox{0.5\ht2-0.5\ht1}{\includegraphics[width = .3\textwidth]{figs/CoolPlate.pdf}}
			\label{fig:cool-plate}
		}
		\hspace{10mm}
		\subfigure[]{
			\includegraphics[width=0.55\textwidth]{figs/CoolSOI.pdf}
			\label{fig:cool-SOI}
			}\end{center} \begin{center}
		\subfigure[]{
			\raisebox{0.5\ht4-0.5\ht3}{\includegraphics[width = .3\textwidth]{figs/CrossSecCooling.pdf}}
			\label{fig:cool-plate-sec}
		}
		\hspace{10mm}     
		\subfigure[]{
			\hspace{0.075\textwidth}
			\includegraphics[width=0.4\textwidth]{figs/AcousticsCooling.pdf}
			\hspace{0.075\textwidth}
			\label{fig:cool-waf}
		}
	\end{center}
	\caption{\protect\subref{fig:cool-plate} Cooling plate wafer etched with micro-channels and closed with glass. \protect\subref{fig:cool-SOI} Schematics of the silicon wafer stack forming a cooling plate. The blue layers indicate the oxidised side of the wafers. See text for explanations. \protect\subref{fig:cool-plate-sec} Scanning electron microscope image of the cross-section of a \vu{150}{\mum} thick prototype cooling plate produced and characterised to evaluate the limit of the technology. \protect\subref{fig:cool-waf} Acoustics imaging of a cooling plate wafer. Two cooling plates are placed in one wafer and are surrounded by testing devices. Each plate contains two cooling circuits.}
	\label{fig:cool-proc}
\end{figure}

The designed cooling device consists of a \vu{210}{\mum} thick \vu{70\times80}{\sqmm} silicon plate fabricated by bonding silicon wafers together, one of which etched to form 150 channels with a cross section of \vu{200\times70}{\sqmum} and arranged in two circuits as shown on~\figs{fig:cool-plate,fig:cool-plate-sec,fig:cool-waf}. Such a device stands out from regular micro-channel designs~\cite{hep-tec_TemizEtAl_2015} by the outstanding large area it covers.
					
Initially, the cooling plates were fabricated by etching microchannels into a bulk silicon substrate and bonding another bulk wafer on top of it to seal the microchannels. In order to thin the cooling plates, the bonded stack was etched in the central region on one or both sides using deep reactive ion etching. An accelerated etching was observed on the edges of the thinned regions. Thus, the cooling plates produced were thinner at the edges of the thinned region, but thicker in the most central part of it. As a result, a low production yield was reached, and the cooling plates equipping the first modules were thicker than the design value: \vu{280}{\mum} if both sides of the stack were etched and \vu{380}{\mum} if only one side was etched.
					
To overcome these issues, a process based on silicon on insulator (SOI) wafers was developed. The principle is illustrated in \fig{fig:cool-SOI}. Two sets of SOI wafers (1-2) and (3-4) are bonded together such that the resulting stacks feature an oxide inner layer. The two silicon stacks are thinned to the desired thickness. Then channels are formed in stack (1-2) using plasma etching techniques. In parallel, an oxide layer is created on stack (3-4) using a wet oxidation technique. The two stacks are bonded together such that the micro-channels of stack (1-2) are closed by stack (3-4). Inlets and outlets are etched in the assembly. The resulting cooling plate is finally thinned using a plasma etching technique designed to be stopped by the inner oxide layers of the silicon stacks. This technique allows to obtain a planarity and a thickness matching the specifications and the production yield was improved from less than 50\% with the initial process to about 90\% with the SOI-based process.

\begin{figure}[tb]
	\setbox2=\hbox{\includegraphics[width = 0.55\textwidth]{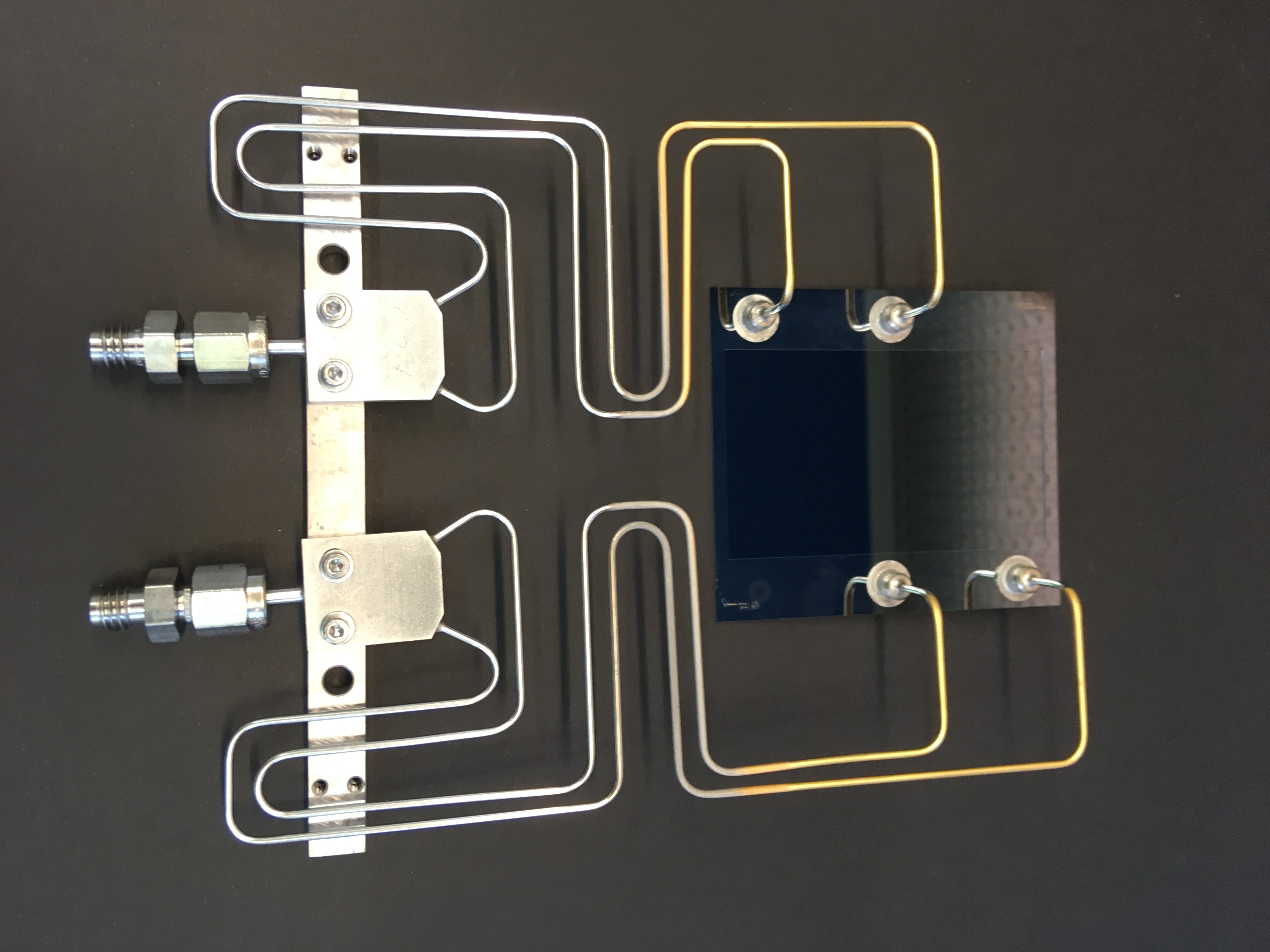}}
	\setbox1=\hbox{\includegraphics[width = 0.3\textwidth]{figs/CoolConnectors.pdf}}
	\begin{center}
		\subfigure[]{
			\raisebox{0.5\ht2-0.5\ht1}{
				\def\svgwidth{0.3\textwidth}
				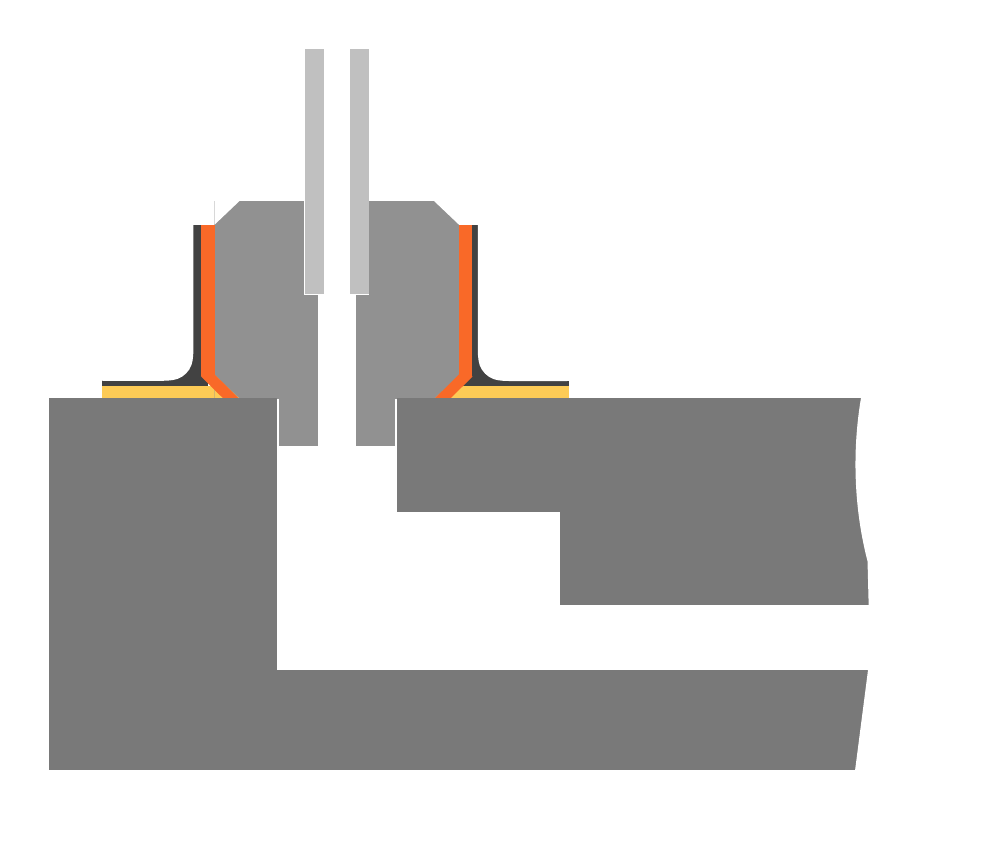
			}
			\label{fig:cool-con}
		}
		\hspace{9mm}
		\subfigure[]{
			\begin{tikzpicture}[x=0.5cm, y=0.5cm]
				\node (image) at (0,0) {\includegraphics[width = 0.55\textwidth]{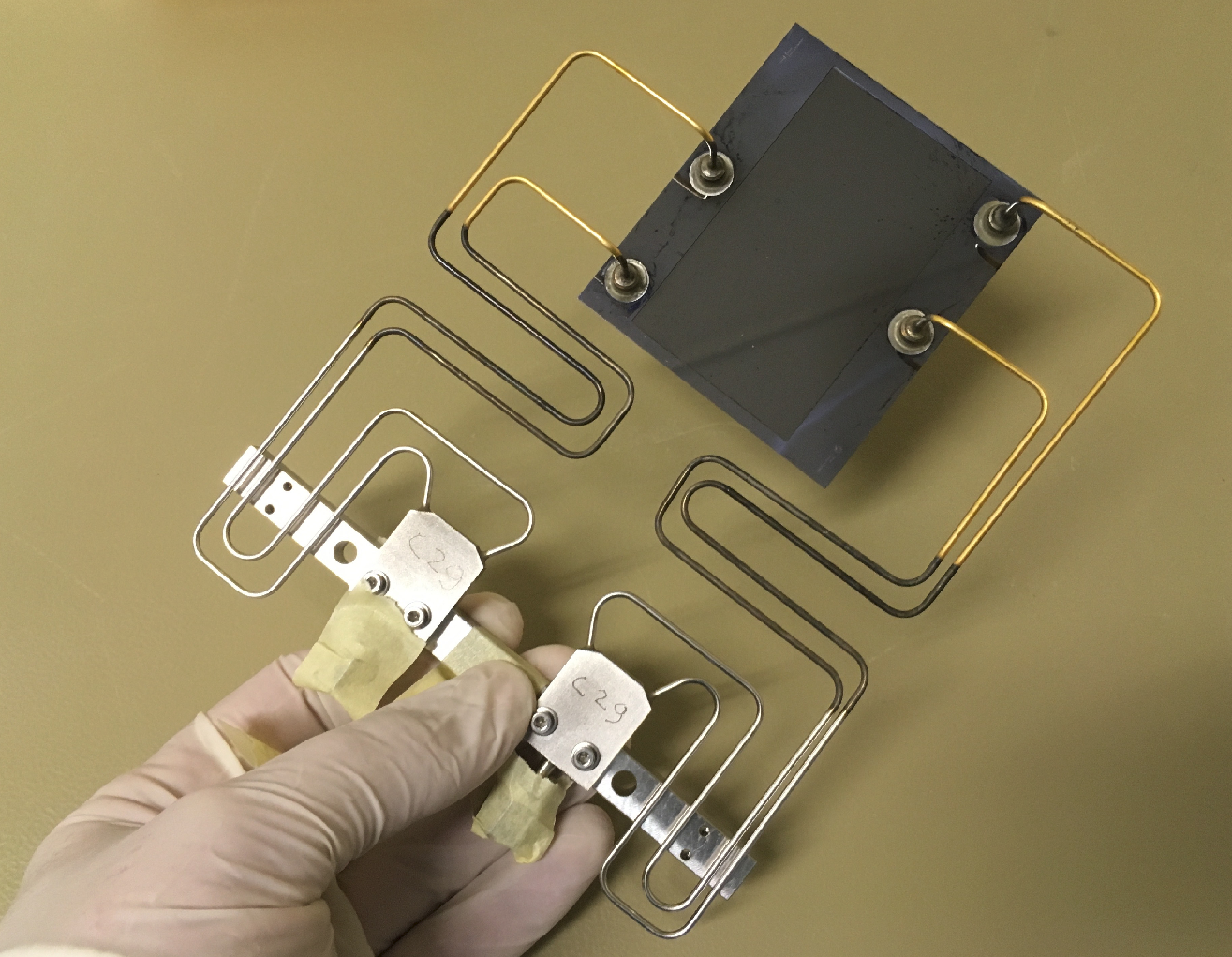}};
				\node[anchor=east] (Capillaries) at (-4,5.7) {Capillaries};
				\draw[->, thick, color=orange!60] (Capillaries) --  (-3,2);
													
				\node[anchor=west] (CP) at (3.7,5.7) {Cooling Plate};
				\draw[->, thick, color=orange!60] (CP) --  (3,4);
													
				\node[anchor=north] (CN) at (5.6,-2.7) {Connector};
				\draw[->, thick, color=orange!60] (CN) --  (4,1.8);
													
				\node[anchor=north west] (MN) at (3.7,-5) {Manifold};
				\draw[->, thick, color=orange!60] (MN) --  (-0.3,-3.2);
													
			\end{tikzpicture}
			\label{fig:cool-cap}
		}
	\end{center}
	\caption{\protect\subref{fig:cool-con} Cross view of the connectors between the plate and the capillaries. \protect\subref{fig:cool-cap} The two cooling plate circuits are connected to capillaries each by one set of inlet and outlet. The capillaries are then brazed on manifolds that are connected to the cooling distribution circuit.}
	\label{fig:cool-con-cap}
\end{figure}

Each of the two plate circuits has an inlet and an outlet, as shown on \fig{fig:cool-cap}. Those openings are metallized such that KOVAR connectors welded with capillaries can be soldered on them as depicted on \fig{fig:cool-con}. The capillaries are routed to exert no stress on the connectors and brazed in manifolds to the cooling distribution circuit.

\subsection{Carrier Board}
					
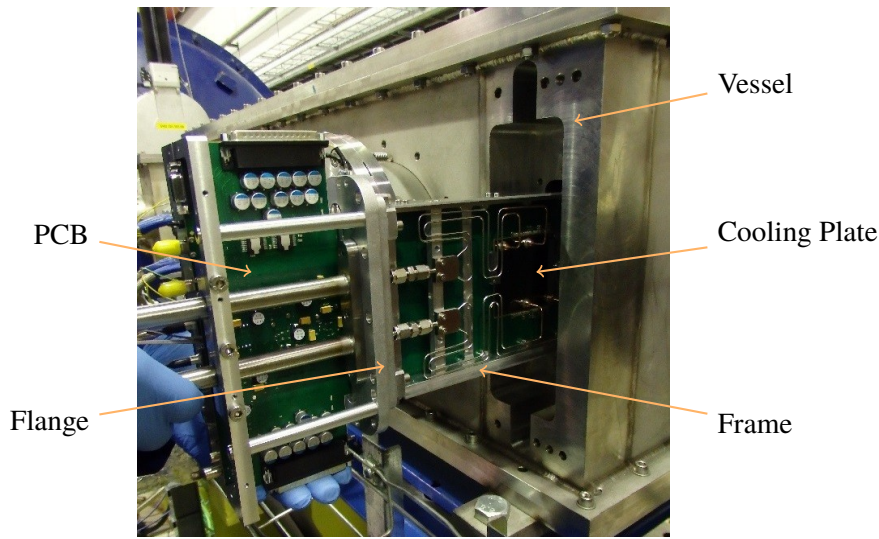
\begin{figure}
	\begin{center}
		\begin{tikzpicture}[x=0.5cm, y=0.5cm]
										
			\node (image) at (0,0) {\includegraphics[width=7cm]{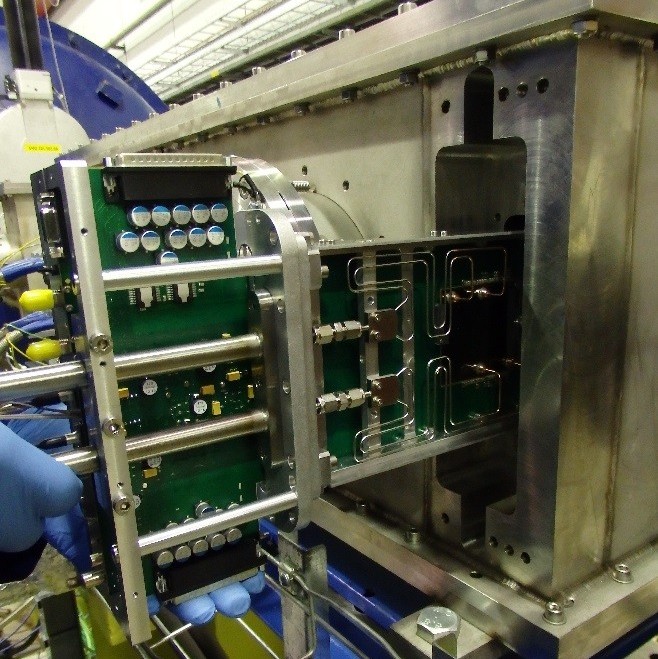}};
										
			\node[anchor=east] (PCB) at (-8,1) {PCB};
			\draw[->, thick, color=orange!60] (PCB) --  (-4,0);
										
			\node[anchor=west, align=left] (CP) at (8,1) {Cooling Plate};
			\draw[->, thick, color=orange!60] (CP) --  (3.5,0);
										
			\node[anchor=west, align=left] (FR) at (8,-4) {Frame};
			\draw[->, thick, color=orange!60] (FR) --  (2,-2.6);  
										
			\node[anchor=east] (FL) at (-8,-4) {Flange};
			\draw[->, thick, color=orange!60] (FL) --  (-0.5,-2.5);  
										
			\node[anchor=west] (VS) at (8,5) {Vessel};
			\draw[->, thick, color=orange!60] (VS) --  (4.5,4); 
										
		\end{tikzpicture}
		\caption{Photograph of a GTK module just before being inserted in the third vacuum vessel. The carrier board is equipped with the assembly made of the HPD and cooling plate and mounted in a frame and a flange acting as vacuum feed-through.}
		\label{fig:carrier}
	\end{center}
\end{figure}

The assembly made of the HPD and the cooling plate is hosted in the carrier board which interfaces it to the systems installed outside the vacuum vessel. Hence, all services to the detector have to pass through the carrier, resulting in a dense and complex design. 
The carrier board is a T-shaped 14-layer printed circuit board (PCB) with a rectangular countersink to receive the assembly made of the HPD and the cooling plate. The board is sealed in an aluminum frame which supports the capillaries manifolds and is equipped with vacuum tight joints. As such, the frame and board, shown on \fig{fig:carrier}, act as vacuum feed-through.
					
\begin{figure}
	\begin{center}
		\begin{tikzpicture}[x=0.45cm, y=0.52cm]
\scriptsize
 \def\myshift#1{\raisebox{1ex}}
 \def\nyshift#1{\raisebox{-2ex}}

\def\T{12}

\node[rectangle,fill=black!10,draw, thick, minimum height=2em, minimum width=5em, align=center, anchor=west, double copy shadow={shadow xshift=0.2em,shadow yshift=-0.15em,draw=black!20,fill=black!10}, label=right:{\footnotesize$\times 10$}] (TDCPix) at (16,11) {TDCPix};

\node[anchor=west](clkDig) at (-2,\T+4.3){Clock Digital};
\node[device] (clkDigR)  at (3,\T+4){Receiver \\ \tiny{OPT1355}};
\node[device] (clkDigFA) at (9,\T+4) {Fanout~1x12\\\tiny{SY89113U}};
\draw[->, thick] (clkDigR) -- (clkDigFA);
\draw[->, thick] (clkDigFA) .. controls ([xshift=+2cm]clkDigFA) and ([yshift=+2cm, ]TDCPix) .. (TDCPix);
\draw[->, thick] (0,\T+4) -- (clkDigR);

\node[anchor=west](clkDll) at (-2,\T+2.3){Clock DLL};
\node[device] (clkDllR) at (3,\T+2){Receiver \\ \tiny{OPT1355}};
\node[device] (clkDllFA)at (9,\T+2) {Fanout~1x12\\\tiny{SY89113U}};
\draw[->, thick] (clkDllR) -- (clkDllFA);
\draw[->, thick] (clkDllFA) .. controls ([xshift=+2.3cm]clkDllFA) and ([yshift=1cm,xshift=-0.5cm ]TDCPix) ..  ($(TDCPix.north)+(-0.5,0)$);
\draw[->, thick] (0,\T+2) -- (clkDllR);

\node[anchor=west](TestPulse) at (-2,\T+0.3){Test Pulse};
\node[device] (DiodeR) at (3,\T+0){Photodiode\\\tiny{FTPDA-R155}};
\node[device] (DiodeFA)at (9,\T+0) {Fanout~1x12\\\tiny{SY89113U}};
\draw[->, thick] (DiodeR) -- (DiodeFA);
\draw[->, thick] (DiodeFA.east) .. controls ([xshift=+0.1cm]DiodeFA.east) and ([yshift=0.5cm,xshift=-0.4cm ]TDCPix.west) ..  ($(TDCPix.west)+(0,0.4)$);
\draw[->, thick] (0,\T+0) -- (DiodeR);

\node[anchor=west,align=left](RstFC) at (-2,\T-1.9){Reset Frame\\Counter};
\node[device] (RstDiodeR) at (3,\T-2){Photodiode\\\tiny{FTPDA-R155}};
\node[device] (RstDiodeFA)at (9,\T-2) {Fanout~1x12\\\tiny{SY89113U}};
\draw[->, thick] (RstDiodeR) -- (RstDiodeFA);
\draw[->, thick] (RstDiodeFA.east) .. controls ([xshift=+0.1cm]RstDiodeFA.east) and ([yshift=-0.5cm,xshift=-0.4cm ]TDCPix.west) ..  ($(TDCPix.west)+(0,-0.4)$);
\draw[->, thick] (0,\T-2) -- (RstDiodeR);

\node[anchor=west](RstGlb) at (-2,\T-3.7){Global Reset};
\node[device] (GRstDiodeR) at (3,\T-4){Photodiode\\\tiny{FTPDA-R155}};
\node[device] (GRstDiodeFA)at (9,\T-4) {Fanout~1x12\\\tiny{SY89113U}};
\draw[->, thick] (GRstDiodeR) -- (GRstDiodeFA);
\draw[->, thick] (GRstDiodeFA) .. controls ([xshift=+2.3cm]GRstDiodeFA) and ([yshift=-1cm,xshift=-0.5cm ]TDCPix) ..  ($(TDCPix.south)+(-0.5,0)$);
\draw[->, thick] (0,\T-4) -- (GRstDiodeR);

\node[anchor=west](SerConf) at (-2,\T-5.7){Configuration};
\node[device] (SConfR) at (3,\T-6) {10Gb/s Rx\\\tiny{AFBR821FH12}};
\draw[->, thick] (SConfR) .. controls ([xshift=+4cm]SConfR) and ([yshift=-2cm, ]TDCPix) .. (TDCPix);
\draw[->, thick] (0,\T-6) --  (SConfR);

\node[rectangle, rounded corners, draw, thick, minimum width=6em, align=center,] (PS)  at (18,5) {Power Supplies
\\ \tiny \vu{1.30}{V}-\vu{0.31}{A} [$\times 10$] 
\\ \tiny \vu{1.35}{V}-\vu{2.60}{A}  [$\times 10$] 
\\ \tiny \vu{2.5}{V}-\vu{2.93}{A}  [$\times 1$] 
\\ \tiny \vu{3.3}{V}-\vu{0.74}{A}  [$\times 1$] 
\\ \tiny \vu{5.0}{V}-\vu{0.33}{A}  [$\times 1$] 
};
\draw[->, thick] ($(PS.north)+(0.1,0)$) --  ($(TDCPix.south)+(0.5,0)$);

\node[anchor=west,align=left] (interlock)at (28,\T+2) {Temperature\\Interlock};
\draw[->, thick] ($(TDCPix.north)+(0.5,0)$) .. controls ([yshift=1cm,xshift=0.5cm ]TDCPix) and ([xshift=-2.3cm]interlock)  ..  (interlock);

\node[device,double copy shadow={shadow xshift=0.2em,shadow yshift=-0.15em,draw=black!20}, label=above right:{\footnotesize$\times 4$}] (MiniPods) at (22,\T+0) {10Gb/s Tx\\\tiny{AFBR-811FH1Z}};
\draw[->, thick] ($(TDCPix.east)+(0,0.4)$) .. controls ([yshift=+0.5cm,xshift=0.4cm ]TDCPix.east) and ([xshift=-0.4cm ]MiniPods.west).. (MiniPods.west) ;

\node[anchor=west,align=left](DataOut) at (interlock.west|-MiniPods.east){Data Out};
\draw[->,thick,postaction={decorate,decoration={text along path,text align=right,text={| \myshift|{$\times$}40}}}]  (MiniPods.east) -- (DataOut);
      
\node[device] (MiniPodCfg) at (22,\T-2) {10Gb/s Tx\\\tiny{AFBR-811FH1Z}};
\draw[->, thick] ($(TDCPix.east)+(0,-0.4)$) .. controls ([yshift=-0.5cm,xshift=0.4cm ]TDCPix.east) and ([xshift=-0.4cm ]MiniPodCfg.west).. (MiniPodCfg.west) ;

\node[anchor=west,align=left](CfgOut) at (interlock.west|-MiniPodCfg.east){Configuration};
\draw[->,thick,postaction={decorate,decoration={text along path,text align=right,text={| \myshift|{$\times$}10}}}]  (MiniPodCfg.east) -- (CfgOut);

\end{tikzpicture}
		\caption{Simplified schematics of the carrier electronics.}
		\label{fig:carrier-elec}
	\end{center}\end{figure}
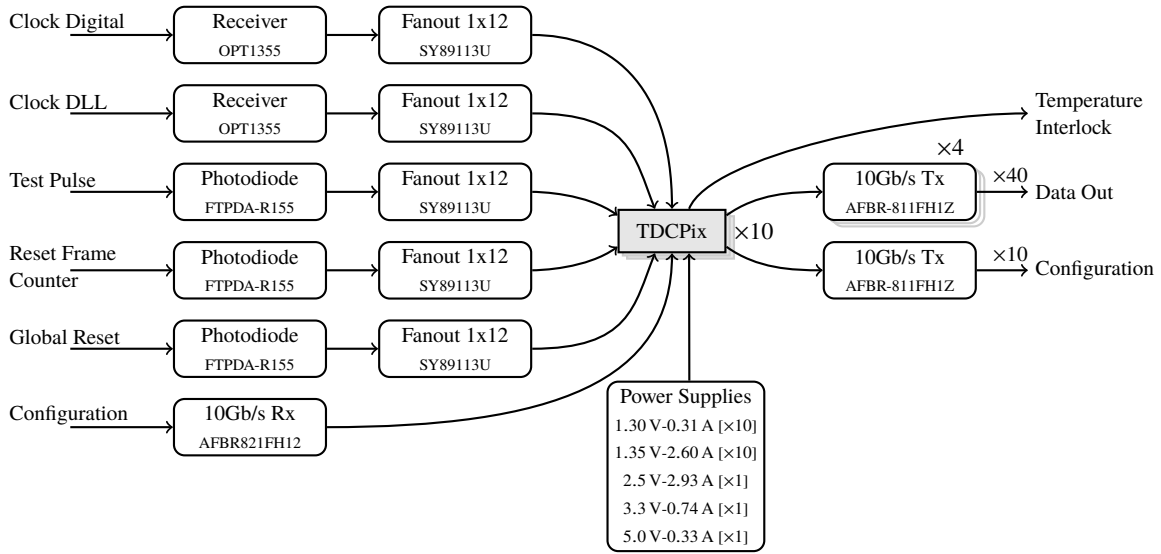

The assembly made of the HPD and cooling plate is clamped from below in the countersink. The planarity around this opening is therefore required to be better than \vu{100}{\mum}. The top side of the opening is machined to form a staircase equipped with 1450 bonding pads spaced by \vu{73}{\mum} onto which the chips are wire bonded. Power, configuration instructions, state information, interlock signals and hit data are transferred through those bonds to the board in which they are routed on \vu{\sim30}{cm} long strips outside the vacuum vessel. The board is equipped with opto-electrical devices connected with optical fibres to the readout boards installed on surface. A simplified schematics of the carrier electronics is shown in \fig{fig:carrier-elec}. 
						
	\subsection{Readout System}
								
	The GTK readout system is the interface between the GTK stations and the rest of the NA62 experiment. The system provides an NA62 wide synchronised clock to the stations, it transfers, the configuration and control commands from the NA62 run control to the stations, and it receives the station data and extracts from them the hits matching the NA62 trigger requests.  
	The latter task is the most challenging as all the data produced by the TDCPix ASICs are sent out to the readout system in a continuous unfiltered stream. As each chip sends data over four \vu{3.2}{Gb/s} serialisers, the data rate can reach up to \vu{128}{Gb/s} per station.
								
	The GTK readout system consists of three sets, one per station, of ten custom electronics boards, serving one TDCPix each. Each set is connected to two computers. The boards comprise a motherboard onto which a daughter card is plugged. The mother board is equipped with:
	\begin{itemize}
		\item a field-programmable gate array (FPGA, Altera Stratix GX110) with two \vu{1}{GB} DDR2 SDRAM, mostly dedicated to the trigger matching operations,
		\item four \vu{3.2}{Gb/s} optical links dedicated to the TDCPix data,
		\item one \vu{320}{Mb/s} optical link dedicated to the TDCPix configuration and
		\item two Gigabit Ethernet copper links to communicate with the computers.
	\end{itemize}
								
	The daughter board implements the TTC protocol~\cite{hep-tec_RD12_1998} with which clock and trigger signals are handled. The board is also equipped with a spare Gigabit Ethernet copper link.
								
	The computers are dual CPU servers running Linux. They are equipped with a least \vu{24}{GB} of RAM and two \vu{10}{Gb/s} Ethernet ports. One port is used to send the trigger matched data to the NA62 computer farm. The other one is used to receive the data from five boards multiplexed by an Ethernet switch.

	\begin{figure}[b]
		\begin{center}
			\import{figs/}{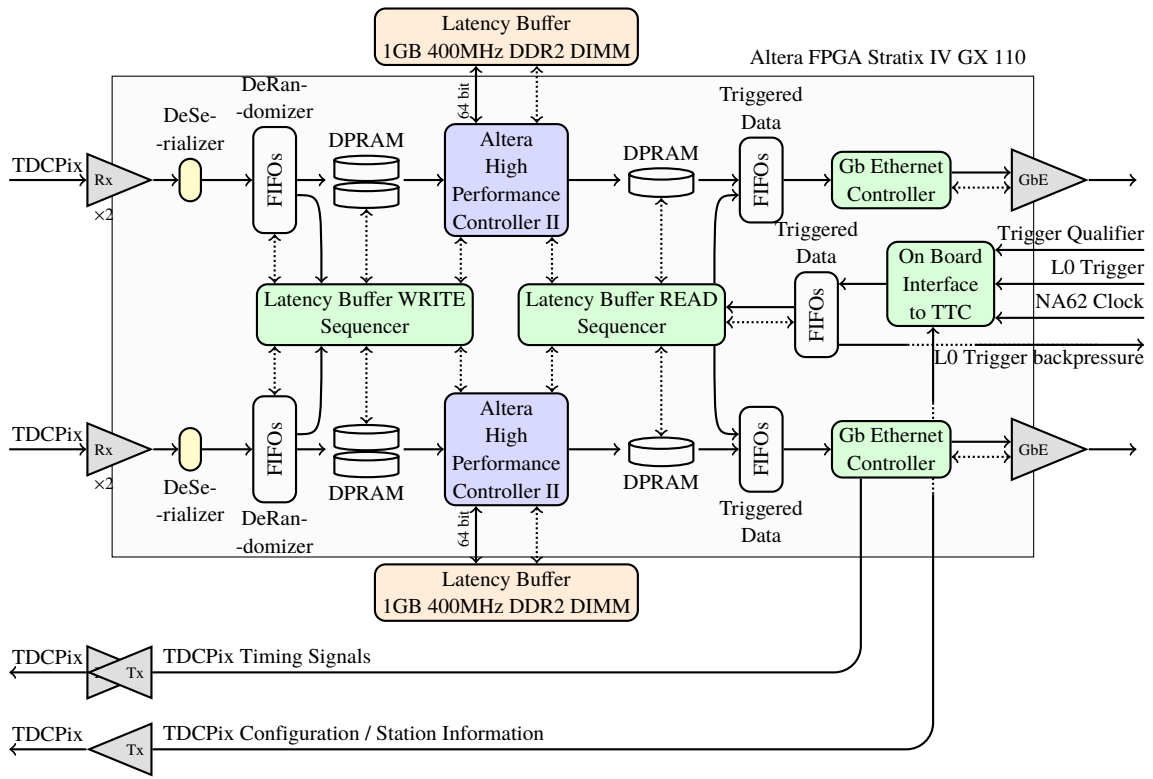}
			\caption{Simplified block diagram of the logic implemented in the FPGA to extract from the GTK data the hits matching a trigger request. See text for explanations.}
			\label{fig:readout}
		\end{center}\end{figure}
										
		The boards receive the NA62 asynchronous L0 trigger requests, with a rate up to \vu{1}{MHz}, extract the corresponding pieces of input data and send them to the six computers. A simplified block diagram of the logic implemented in the FPGA to extract the hits matching a trigger request from the GTK data is shown in \fig{fig:readout}.
		Each GTK readout is logically divided in two identical halves, each one handling the data coming from two of the four serializers of each TDCPix. Incoming data are first of all de-serialized, assembled in a segment of cache DPRAM spanning \vu{6.4}{\mus} and then copied to the DDR2 memory waiting for level-0 trigger decision (the maximum level-0 trigger latency is 1 ms). Two such segments (the even and odd ones) exist and are cycled between them so that while the even one is receiving the data, the odd one is being copied to DDR2 (and vice-versa). To extract the hits matching a trigger request, data are copied from DDR2 memory to a different DPRAM segment, and data in a \vu{75}{ns} time window around the trigger time-stamp are selected. These data are then assembled in UDP packets and sent to the readout servers. Upon asynchronous L1 request, with a rate up to \vu{100}{kHz}, each server assembles the data in one packet and sends it to the NA62 computer farm.

		\subsection{Services}
										
		The operation of the GTK stations relies on the provision of refrigerated coolant, stable electrical power and vacuum.
										
		\subsubsection*{Cooling Plant}
		A cooling plant was assembled to provide the three stations with refrigerated perfluorohexane - $\rm{C_6F_{14}}$. To reduce the exposition to radiation, the plant is located \vu{10}{m} away from the beam line. The coolant is provided to the stations through three lines of  \vu{10}{m} to \vu{20}{m} lengths. The operating detector pressure is \vu{3}{bars} with a coolant flow of \vu{3}{g/s}. The coolant temperature at the detector inlet is adjusted with heat exchangers installed just before the stations and fed in $\rm{C_6F_{14}}$ by the cooling plant though dedicated high flow distribution lines.
										
		The detector distribution lines are equipped with pressure regulating analogue valves to limit the input pressure. In addition, robust operating procedures were developed to ensure smooth pressure and temperature ramping at the detector during starting and stopping phases. To avoid risks related to condensation, the cooling plant is interlocked with the detector vacuum. All the programmable logic controller (PLC) electronics are located in a radiation protected area to avoid perturbations due to the accumulated dose.
										
		\subsubsection*{Vacuum}
										
		The detectors are installed in three vacuum vessels operated at a pressure of \vu{\sim 10^{-7}}{mbar}. Each vessel can be isolated from the beam vacuum by valves at the input and the output. The third station vessel hosts as well the CHANTI detector. A coarse vacuum is provided in each vessel by a turbo-molecular pump. The nominal pressure is obtained opening all the valves and using the cryo-pumps installed in the downstream part of the beam pipe.
										
		\subsubsection*{Power}
										
		The detector electronics power, except the sensor high voltage, is provided by radiation hard power supplies installed \vu{10}{m} away from the beam line to reduce the radiation exposition. The power supplies cables are \vu{10}{m} to \vu{20}{m} long. Power cables are thus equipped with sense wires through which the voltage delivered at the detector is read out and adjusted with a feed-back loop. The power supplies are interlocked to the TDCPix over-temperature alarm and to the cooling system.
		The high voltage power supplies are installed in the surface counting room and connected to the detector with a \vu{200}{m} long cable. These power supplies are interlocked to the same state-machines as the low voltage power supplies and also to the vacuum gauge to avoid electrical discharges.

 \section{Detector Performance}
  \subsection{Detectors and Operations}
\begin{figure}[h]
 \begin{center}
      \subfigure[]{
      \includegraphics[width = 7cm]{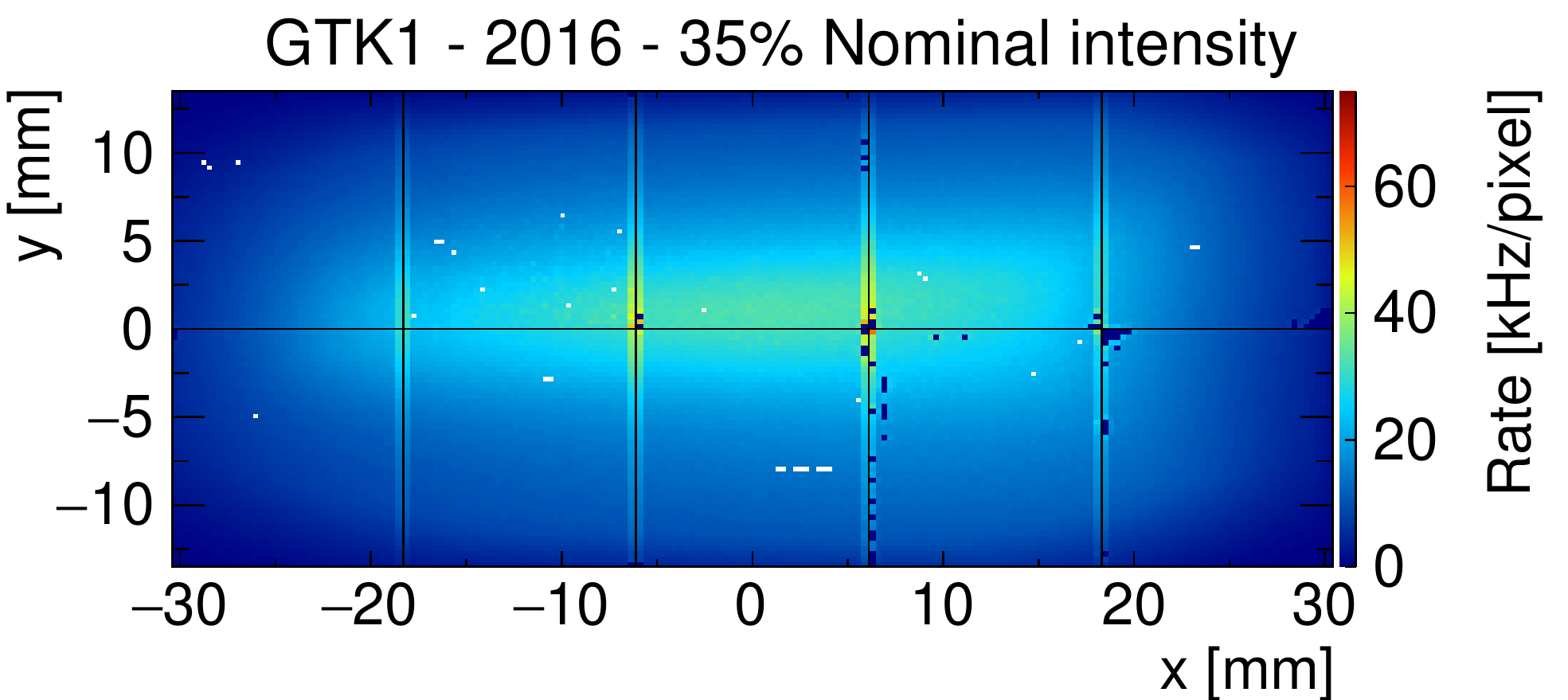}
      \label{fig:hitmap1}
    }
      \subfigure[]{
      \includegraphics[width = 7cm]{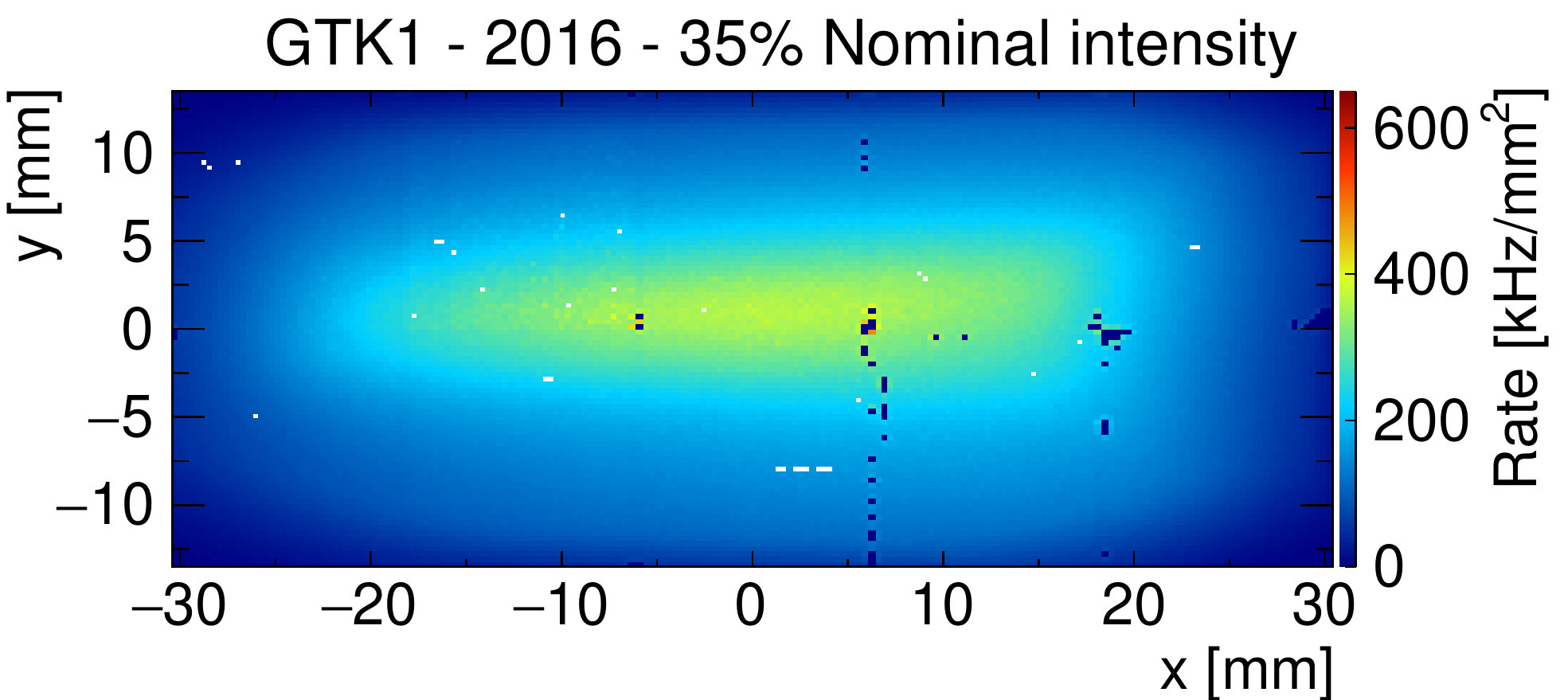}
      \label{fig:hitmap1c}
    }
    \subfigure[]{
    \includegraphics[width= 7cm]{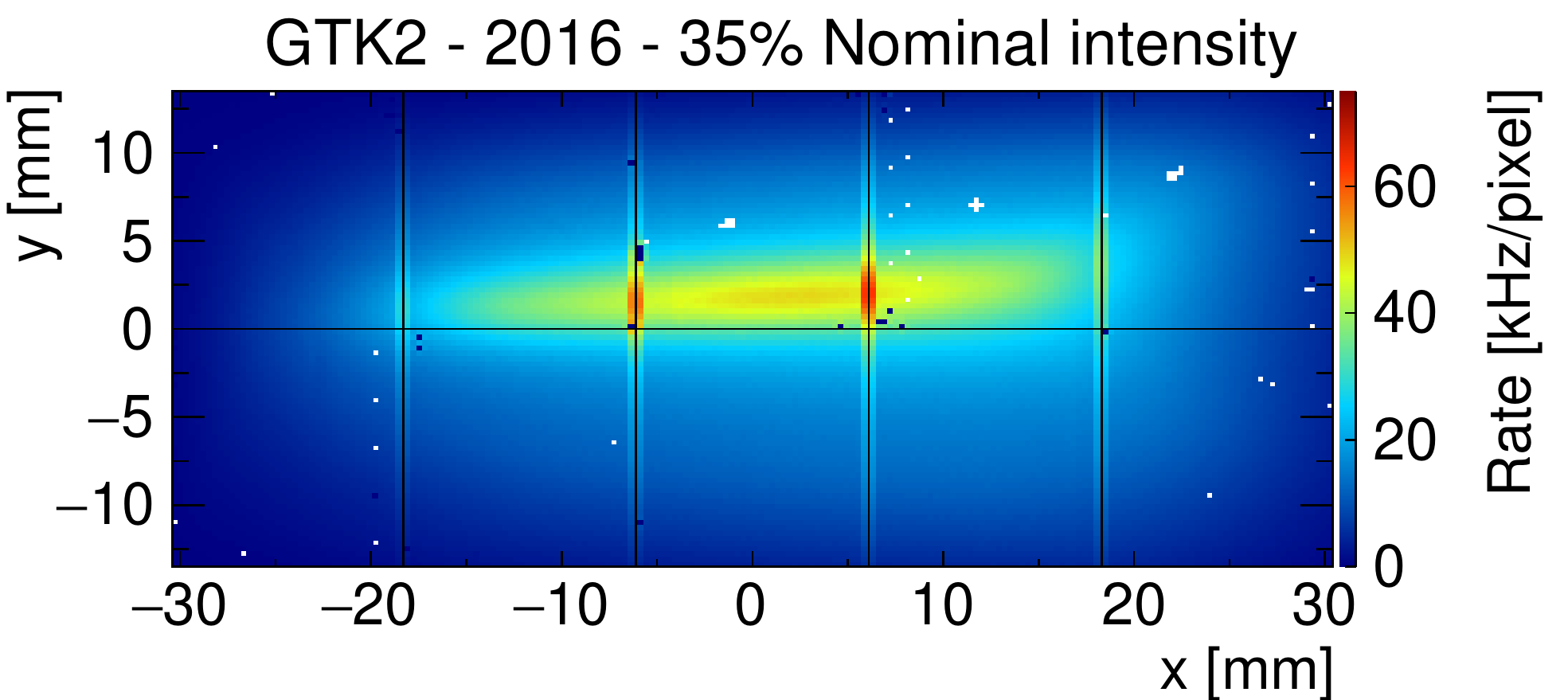}
    \label{fig:hitmap2}
    }
    \subfigure[]{
    \includegraphics[width= 7cm]{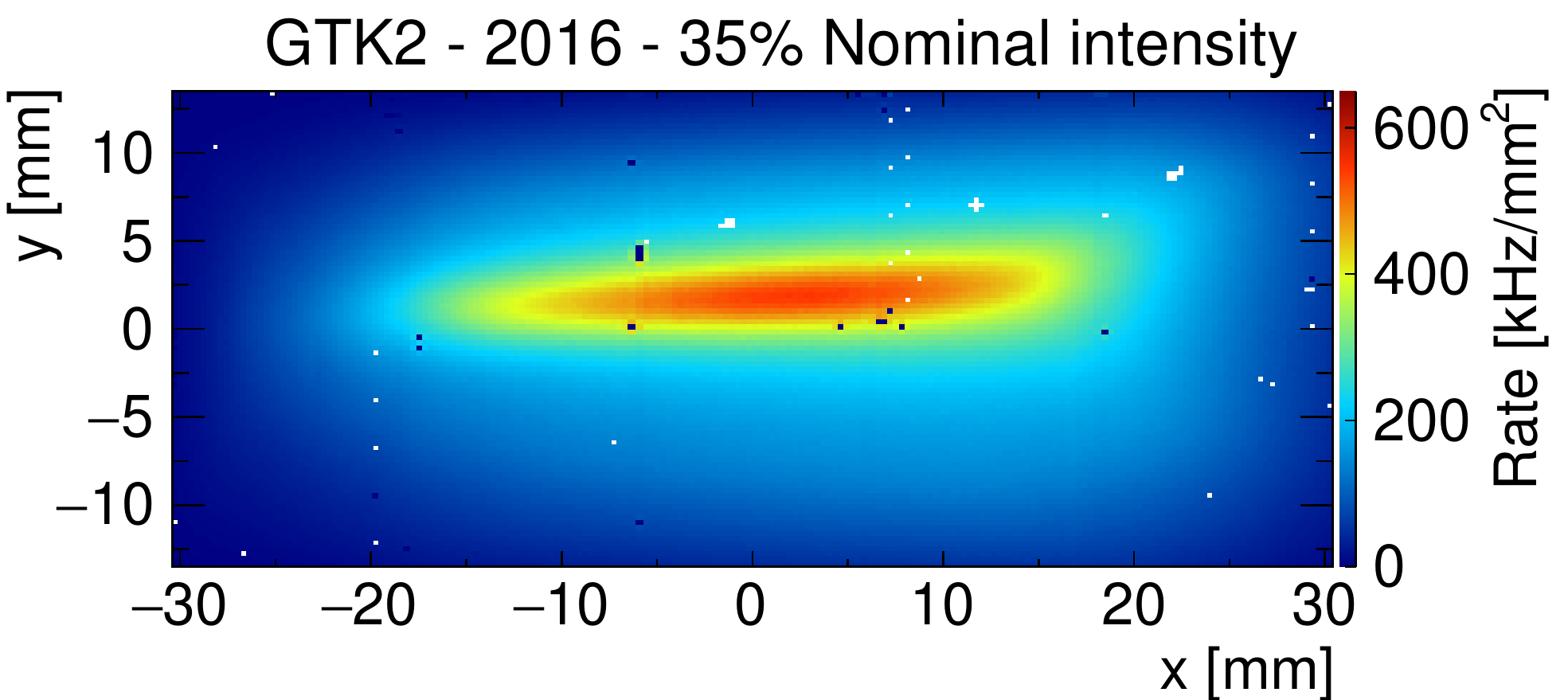}
    \label{fig:hitmap2c}
    }
    \subfigure[]{
    \includegraphics[width= 7cm]{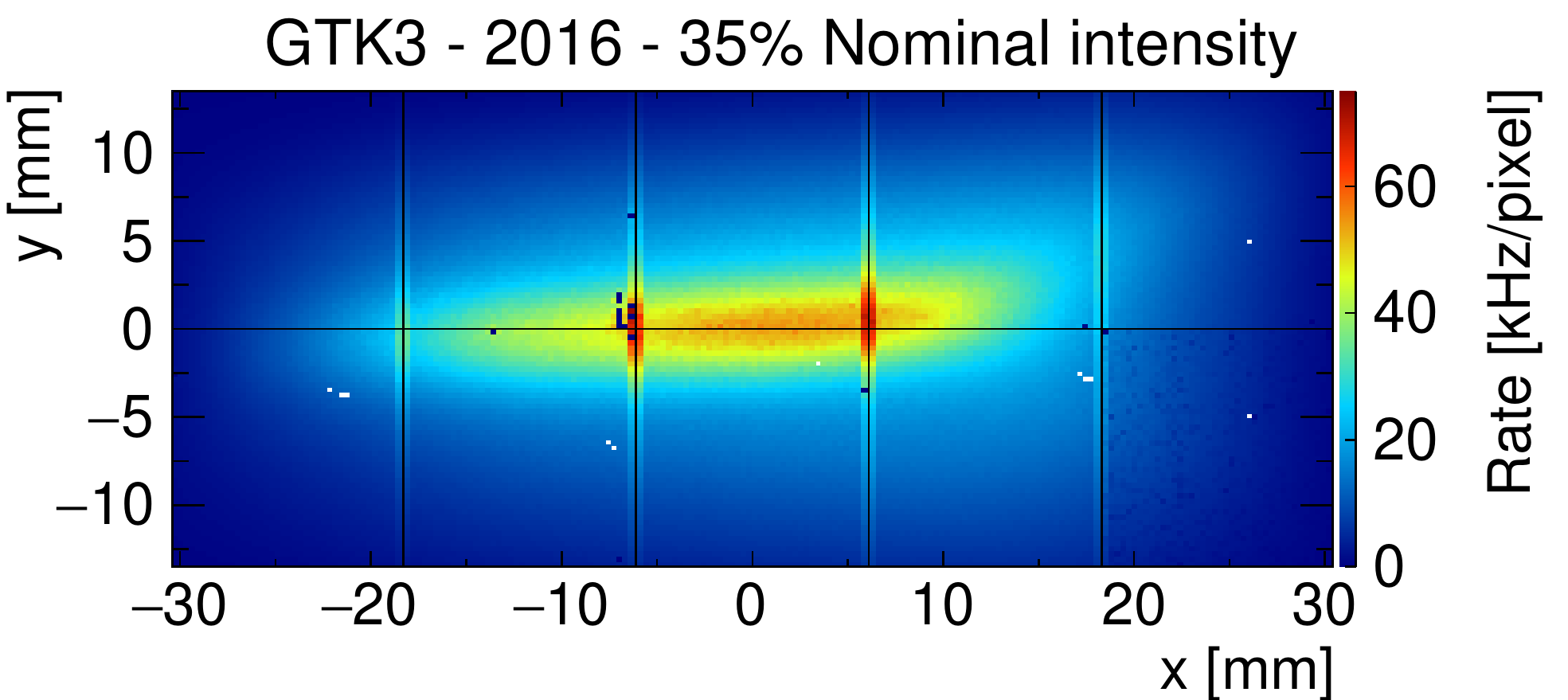}
    \label{fig:hitmap3}
    }
    \subfigure[]{
    \includegraphics[width= 7cm]{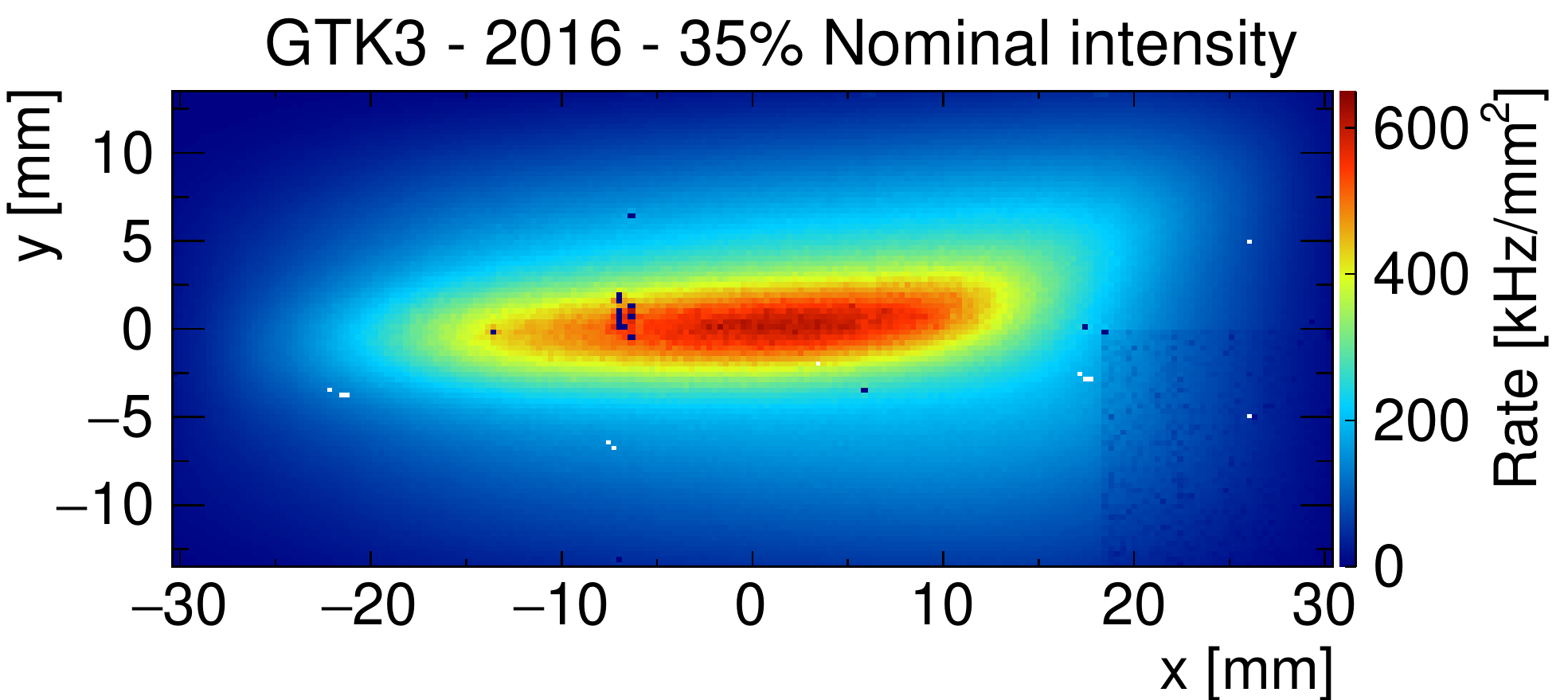}
    \label{fig:hitmap3c}
    }
\end{center}
\caption{Hit rate per pixel (left column in kHz/pixel, right column in kHz/mm$^{2}$) in the three GTK modules recorded in 2016 at around 35\% of nominal beam intensity. In \protect\subref{fig:hitmap1c}, \protect\subref{fig:hitmap2c} and, \protect\subref{fig:hitmap3c} the rates in the elongated pixels is scaled down by 30\% to correspond to a standard pixel. When present, the black lines indicate the readout chips.}
  \label{fig:hitmap}
\end{figure}

 Three stations were constructed for the 2016 data taking operations. They were all equipped with n-in-p sensors and mounted on cooling plates with thickness of \vu{380}{\mum} for GTK1 and GTK2 and and \vu{280}{\mum} for GTK3. The material budget in the beam acceptance is therefore 0.73\% X$_0$ for GTK1 and GTK2 and and 0.62\% X$_0$ for GTK3. The fraction of dead pixels is below 0.5\% which can be seen in the detector hit maps shown in \fig{fig:hitmap}.

 The thresholds of the pixel discriminators were tuned in the laboratory. The expected charge distribution for minimum ionizing particles following a Landau distribution peaking at \vu{2.4}{fC}. The pixel thresholds were trimmed to \vu{0.8}{fC}. The threshold calibration was done using a $\rm{{}^{109}Cd}$ source extrapolating from the \vu{0.98}{fC} and \vu{1.11}{fC} charge depositions of the low energy photons (\vu{22.1}{keV} and \vu{24.9}{keV}) emitted.

 The three GTK stations were  fully commissioned by mid-September 2016 and took data until mid-November 2016. They were then dismounted and stored at \vu{-21}{ \degree C}. The same three stations were installed in the beam line in May 2017 and took data until mid-October 2017.

\begin{figure}[h]
 \subfigure[]{
   \includegraphics[width = .5\textwidth]{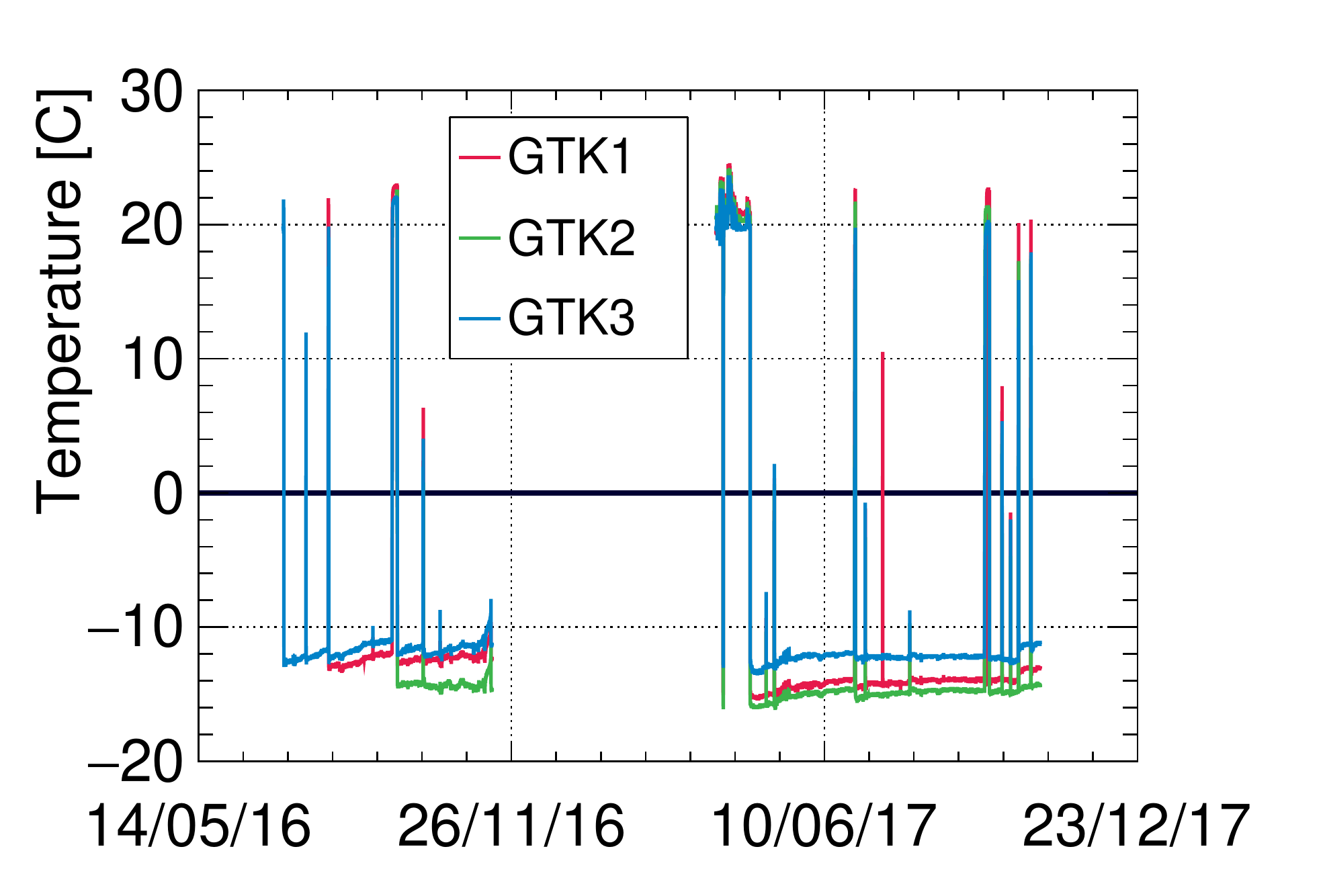}
  \label{fig:temperature}
  }
 \subfigure[]{
  \includegraphics[width = 0.5\textwidth]{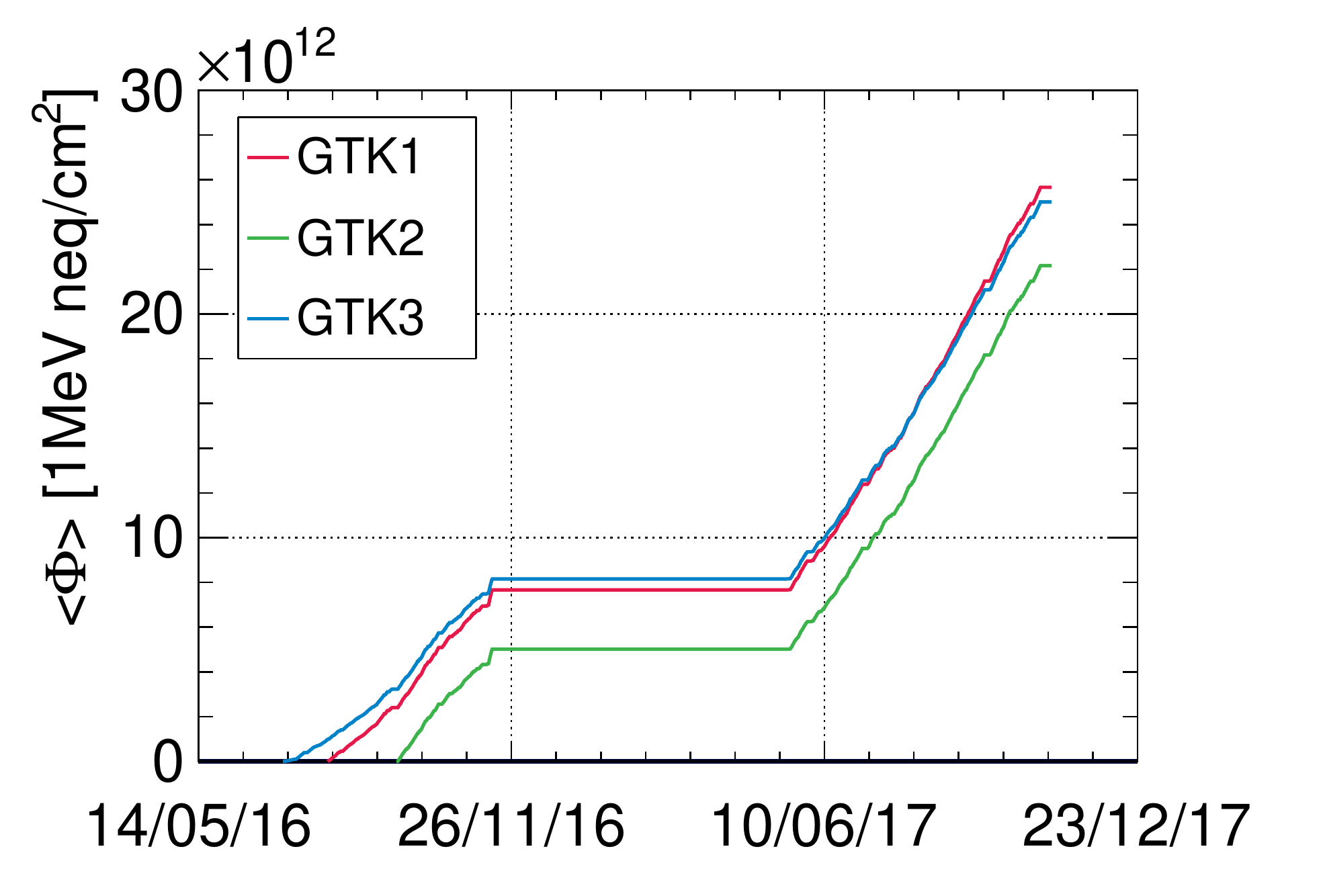}
   \label{fig:fluence}
  }
 \caption{\protect\subref{fig:temperature} Inlet temperature of the three GTK station when installed in the beam line. \protect\subref{fig:fluence} Average integrated fluence of the three GTK stations.}
 \label{fig:conds}
\end{figure}

 During data taking, the stations were operated at a bias voltage of \vu{100}{V} except for few days at the end of the 2016 data taking period when the bias voltage was increased by steps of \vu{50}{V} up to \vu{250}{V}. As shown in \fig{fig:temperature}, the stations were kept cold most of the time and reheated only for short maintenance periods. The beam intensity was around 35\% of the nominal intensity in 2016 and 60-65\% in 2017.  \fig{fig:hitmap} shows the hit rate recorded in each station in 2016. The sensors irradiation was monitored using a proportional chamber operated independently of NA62. The station's average integrated fluence is reported in \fig{fig:fluence}. At the end of the 2017 data taking period, a peak (average) fluence of \vu{1.3~(0.26) \times 10^{14}}{1\,MeV~eq.~n /cm^{2}} was integrated in the stations installed first (GTK1 and GTK3). This fluence corresponds to around 60~days of operation at full beam intensity.

\subsection{Time Resolution}

As explained in Section~\ref{ssec:tdcpix}, the measurement of the hit time coordinate relies on off-line time-walk correction. For every chip of every station, time-walk corrections are derived in \vu{100}{ps} ToT bins between 0 and \vu{30}{ns}. Above this ToT value, the time-walk measurements saturate. 

Moreover, within a chip, pixel output signals are delayed due to the propagation time between the pixels and the EoC depending on the pixel position in the matrix. In addition, the digital signals at the output of the pixel discriminators degrade as they transit in the lines connecting the pixel to the EoC. Hence, the signal delay at the EoC increases with the distance travelled by the signal from the pixel. 
A time correction for each pixel is derived to compensate for these effects.

\begin{figure}[tb]
  \subfigure[]{
    \includegraphics[width = .47\textwidth]{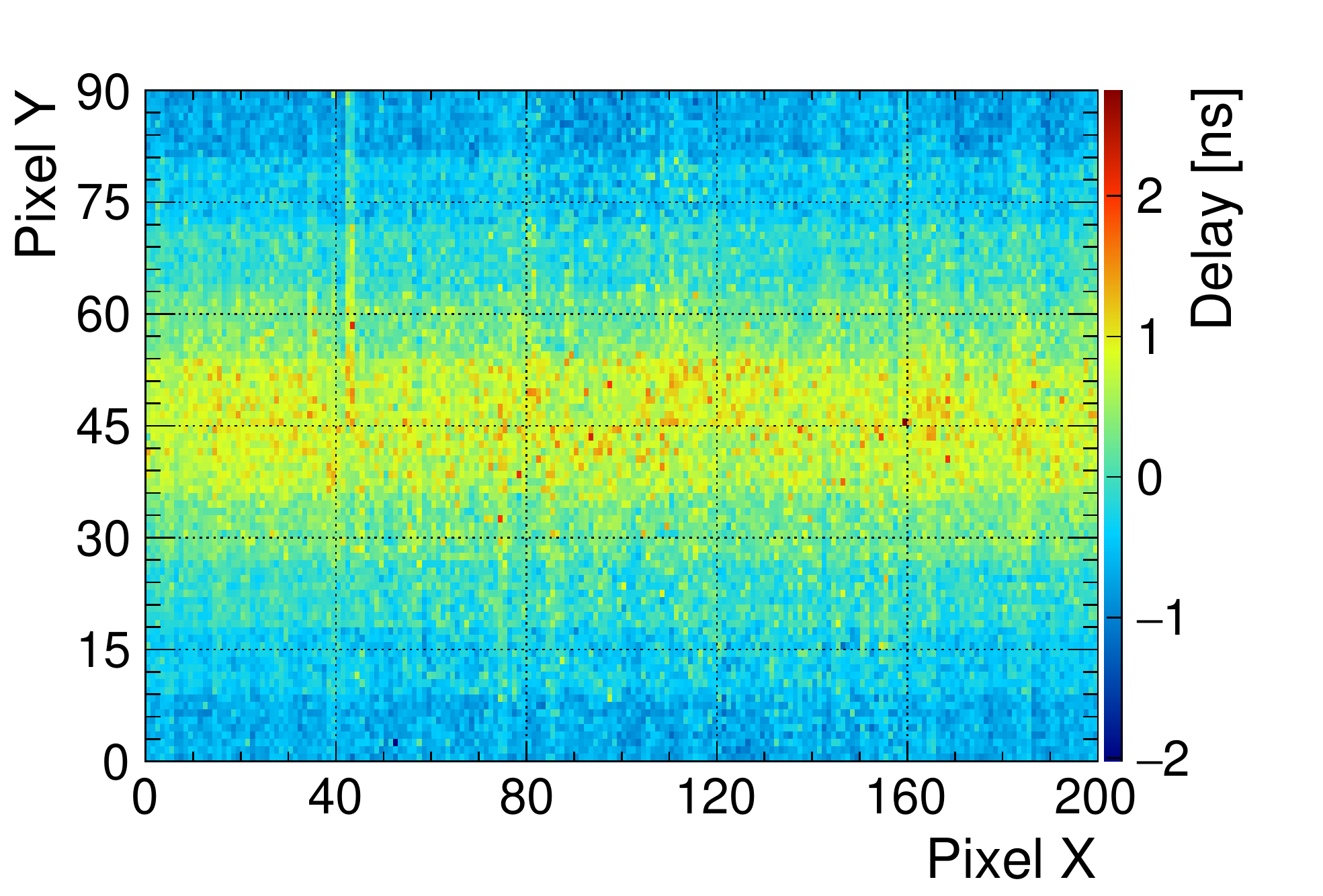}
    \label{fig:gtk-calibration-t0}
  }
  \subfigure[]{
    \includegraphics[width = .47\textwidth]{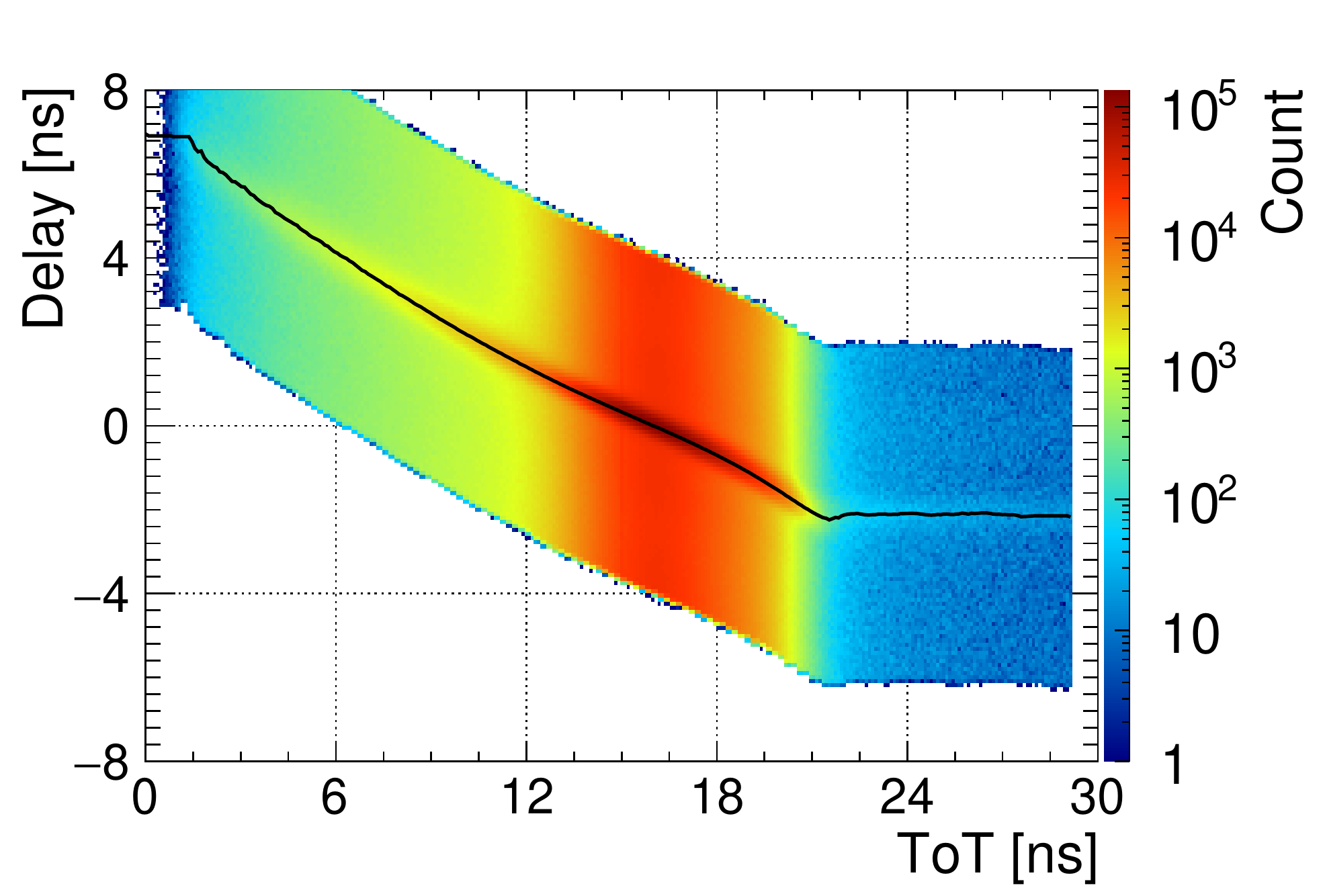}
    \label{fig:gtk-calibration-tw}
  }
  \caption{\protect\subref{fig:gtk-calibration-t0} Station pixel delays as function of pixels coordinates. Delays are offset such that the average delay of the station is null. The pixel delay decreases as function of the distance to the EoC. \protect\subref{fig:gtk-calibration-tw} Chip GTK to KTAG time difference distribution as function of GTK hit ToT, overlaid with fitted time-walk delay (black line).}
  \label{fig:gtk-calibration}
\end{figure}

All the corrections are derived from data using the coincidence of the GTK tracks, formed of three single hit clusters, and a reference time provided by the KTAG or RICH time. The time resolution of these detectors are about \vu{70}{ps} and \vu{100}{ps}, respectively. Example of corrections are shown in \fig{fig:gtk-calibration}.

The time resolution of the detector is measured using samples of \Kpipipi\ collected during 8 to 10 consecutive hours of data taking.
The \Kp, formed by the three pions, is assigned two time coordinates: the average time coordinate of the three pion RICH candidates and the time coordinate of the closest in-time KTAG candidate. Finally, the \Kp\ is associated with a GTK track based on the track kinematics and direction correspondence and no time criteria is applied. 

\begin{figure}[b]
  \subfigure[]{
    \hspace{0.38cm}    \includegraphics[width = .55\textwidth]{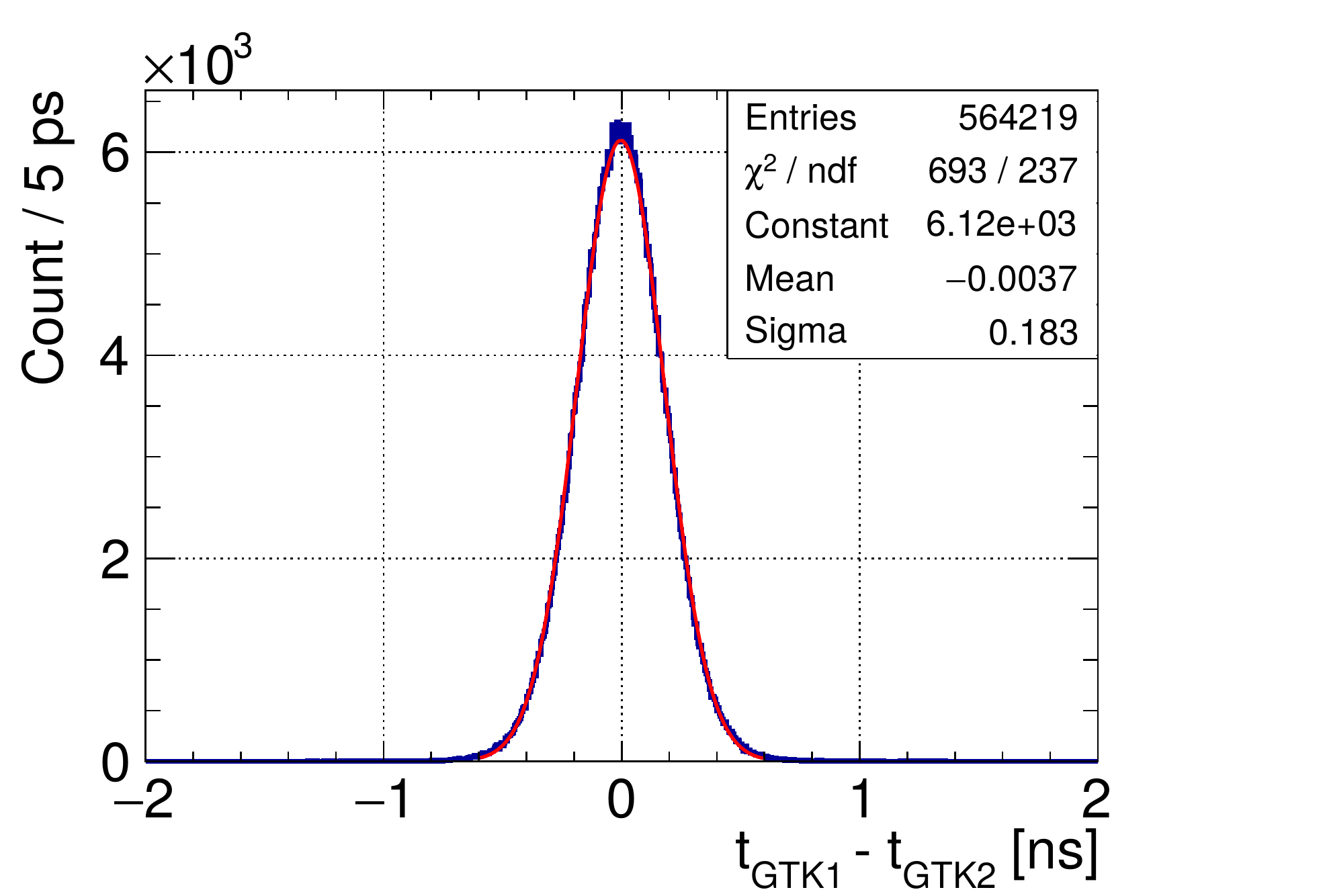}
    \label{fig:standalone-time-diff-plt}
  }
  \subfigure[]{
    \raisebox{2.7cm}{
      \small
      \begin{tabular}{lc}\hline
	\multicolumn{2} {c}{$\sigma$ [ps]}\\\hline
	GTK1-3 	& 181.3\\
	GTK1-2 	& 183.3\\
	GTK2-3 	& 184.7\\\hline
      \end{tabular}\hspace{0.2cm}  
      \begin{tabular}{lc}\hline
	\multicolumn{2} {c}{Resolution [ps]}\\\hline
	GTK1 	& 132.0\\
	GTK2 	& 127.1\\
	GTK3 	& 129.2\\\hline
      \end{tabular}
    }
    \label{fig:standalone-time-diff-tab}
  }
  \caption{\protect\subref{fig:standalone-time-diff-plt} Time difference distribution between hits in the first and second GTK station corresponding to the same \Kp\ track. A Gaussian function is fitted to the distribution (red). \protect\subref{fig:standalone-time-diff-tab} Standard deviation from the fits to the  time difference distributions between the three GTK stations and the corresponding station time resolutions.}
  \label{fig:standalone-time-diff}
\end{figure}

A standalone derivation of the GTK time resolution is performed comparing the three station to station time differences shown in \fig{fig:standalone-time-diff}.
The time resolution values obtained are \vu{132.0}{ps} for GTK1, \vu{127.1}{ps} for GTK2 and \vu{129.2}{ps} for GTK3. Assuming that the track time coordinate is the average of the three hits time coordinates, the time resolution for the track is \vu{74.7}{ps}.

A NA62-wide time resolution is derived fitting Gaussian functions to the time difference distributions between the three GTK stations, the RICH, and the KTAG candidates split by RICH and KTAG candidate hit numbers. The resolutions obtained are compatible with those from the standalone methods and are \vu{135.0}{ps} for GTK1, \vu{128.0}{ps} for GTK2 and \vu{132.4}{ps} for GTK3.

\begin{figure}[!b]
  
  \subfigure[]{
    \includegraphics[width = 0.5\textwidth]{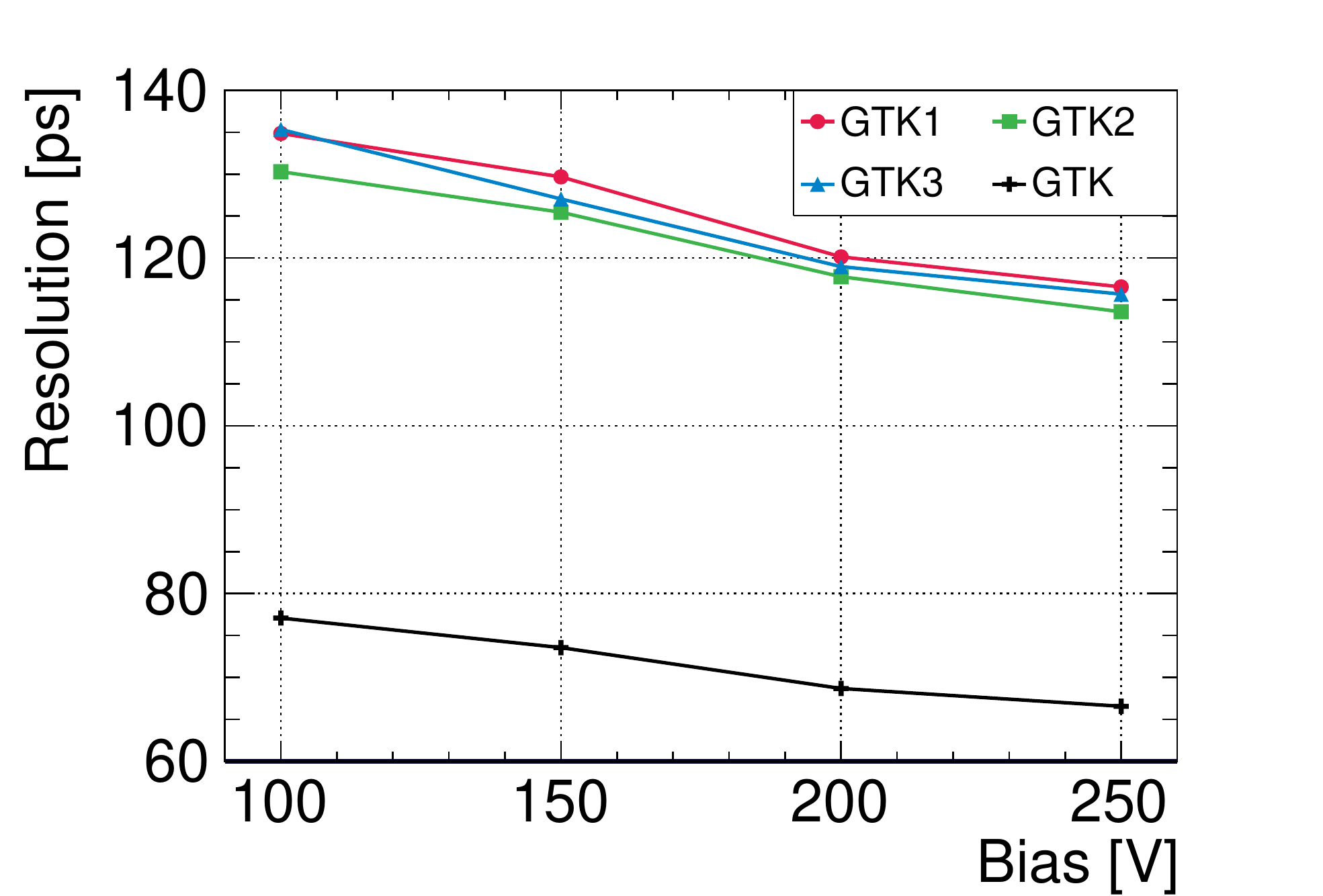}
    \label{fig:reso-bias}
  }%
  \subfigure[]{
    \includegraphics[width = 0.5\textwidth]{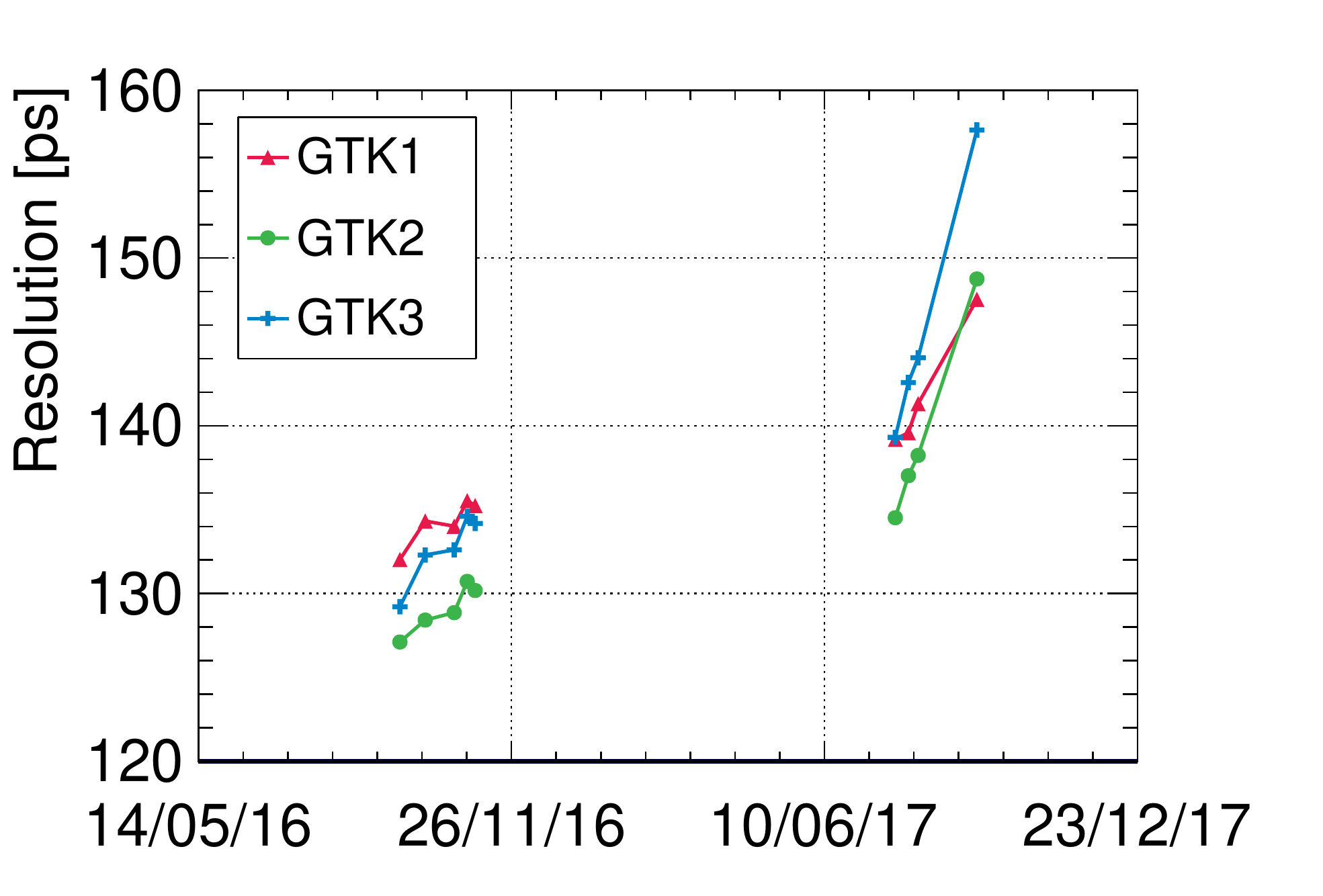}
    \label{fig:reso-t0}
  }\\%
  \subfigure[]{
    \includegraphics[width = 0.5\textwidth]{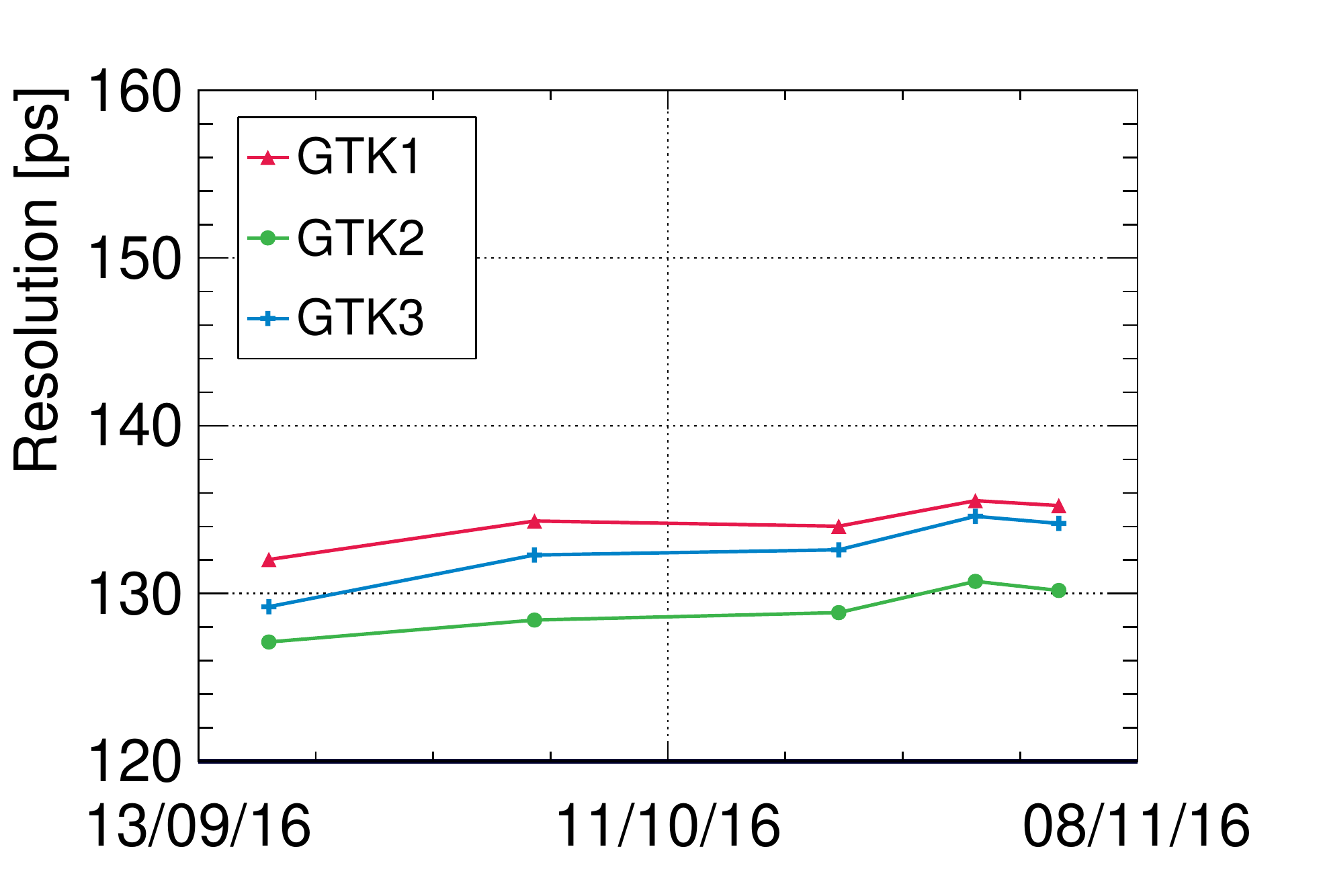}
    \label{fig:reso-t1}
  }%
  \subfigure[]{
    \includegraphics[width = 0.5\textwidth]{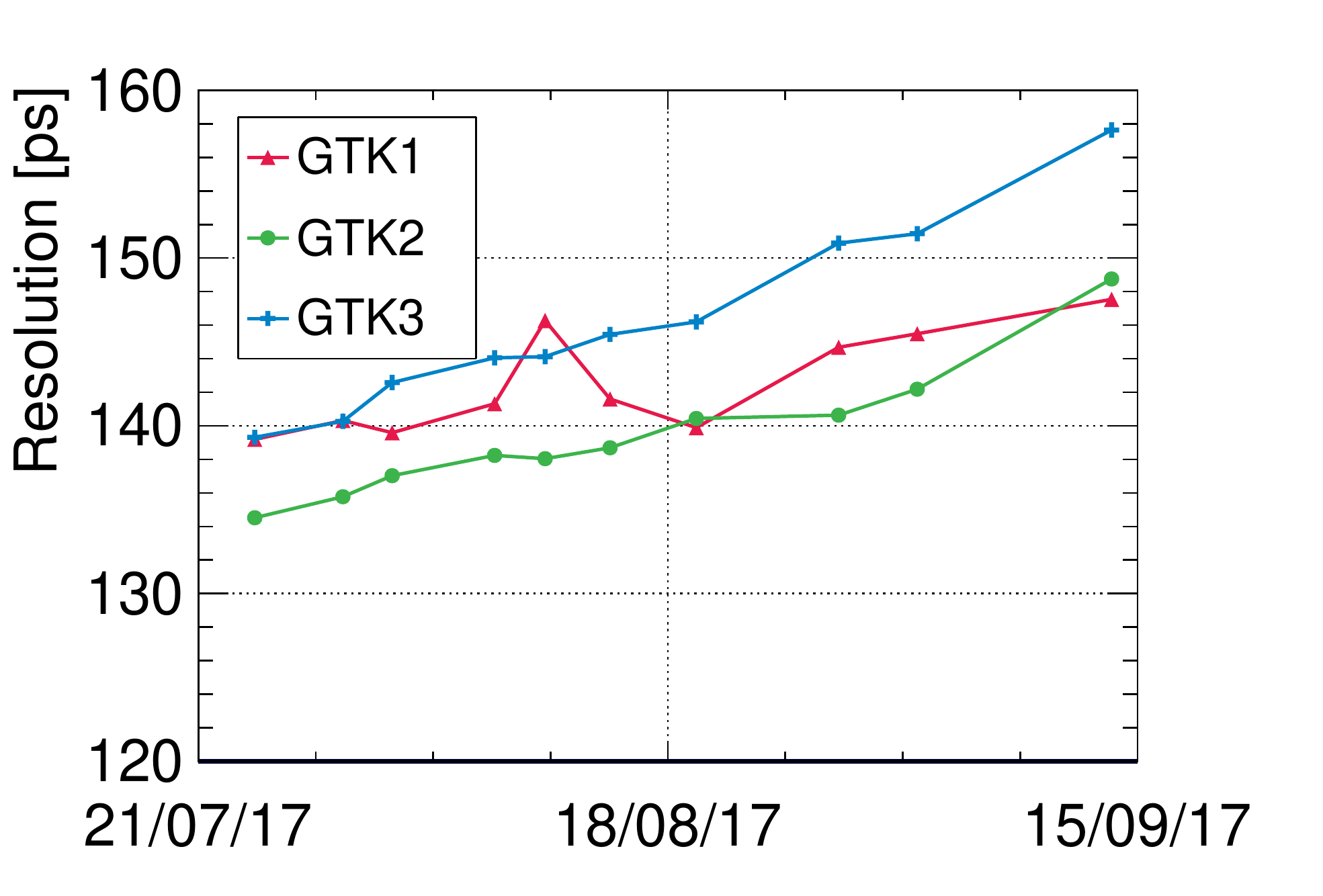}
    \label{fig:reso-t2}
  }%
  \caption{\protect\subref{fig:reso-bias} GTK station time resolution and average time resolution as function of the bias voltage. \protect\subref{fig:reso-t0}--\protect\subref{fig:reso-t2} Evolution of the time resolution over time. Small ticks on the x-axis are spaced by four weeks for \protect\subref{fig:reso-t0} and one week for \protect\subref{fig:reso-t1} and \protect\subref{fig:reso-t2}.}
  \label{fig:resotime}
\end{figure}

The time resolution was also measured at different sensor bias voltages during a ten day period at the end of the 2016 data taking period. The results are reported on \fig{fig:reso-bias} and show a moderate but systematic improvement of the station time resolution of \vu{20}{ps} increasing the sensor bias from \vu{100}{V} to \vu{250}{V}. \fig{fig:reso-bias} also shows that for a bias voltage of 250 V a single station time resolution of 115 ps and a track resolution of 65 ps was achieved.

The time resolution was monitored over time as shown in \fig{fig:reso-t0}--\subref{fig:reso-t2}. The resolution degrades over time and reaches \vu{146-158}{ps} at the end of the 2017 data taking period which is still better than the nominal specifications. Such a degradation was an expected effect of the detector irradiation, however no study rules out other concurrent factors.

\subsection{Kinematics Performances and Material Budget}

The kinematics performances of the GTK are of paramount relevance to reduce the  background from \Kpipizero\ for the \Kpinn\ branching ratio measurement as explained in Section~\ref{sec:intro}. One of the key elements controlling these performances is the station's material budget as shown in \fig{fig:GTKExpectedKineReso}. Moreover, the material budget of the last GTK station is crucial to reduce the background arising from beam particles inelastically scattered in this station. Solid estimates of this quantity are thus needed.

A first estimate is obtained by measuring the thickness of the station sensor, chips and cooling plate. The thickness found correspond to a material budget of 0.73\% X$_0$ for GTK1 and GTK2 and 0.62\% X$_0$ for GTK3. These results can be cross checked for GTK2 using data by measuring the residual along the horizontal axis between the hit position in this station and the position extrapolated based on the hits in the two other stations. The results found are compatible within 10\% with those based on the thickness measurement.

\begin{figure}[bth]
  \begin{center}
    \includegraphics[width = .7\textwidth]{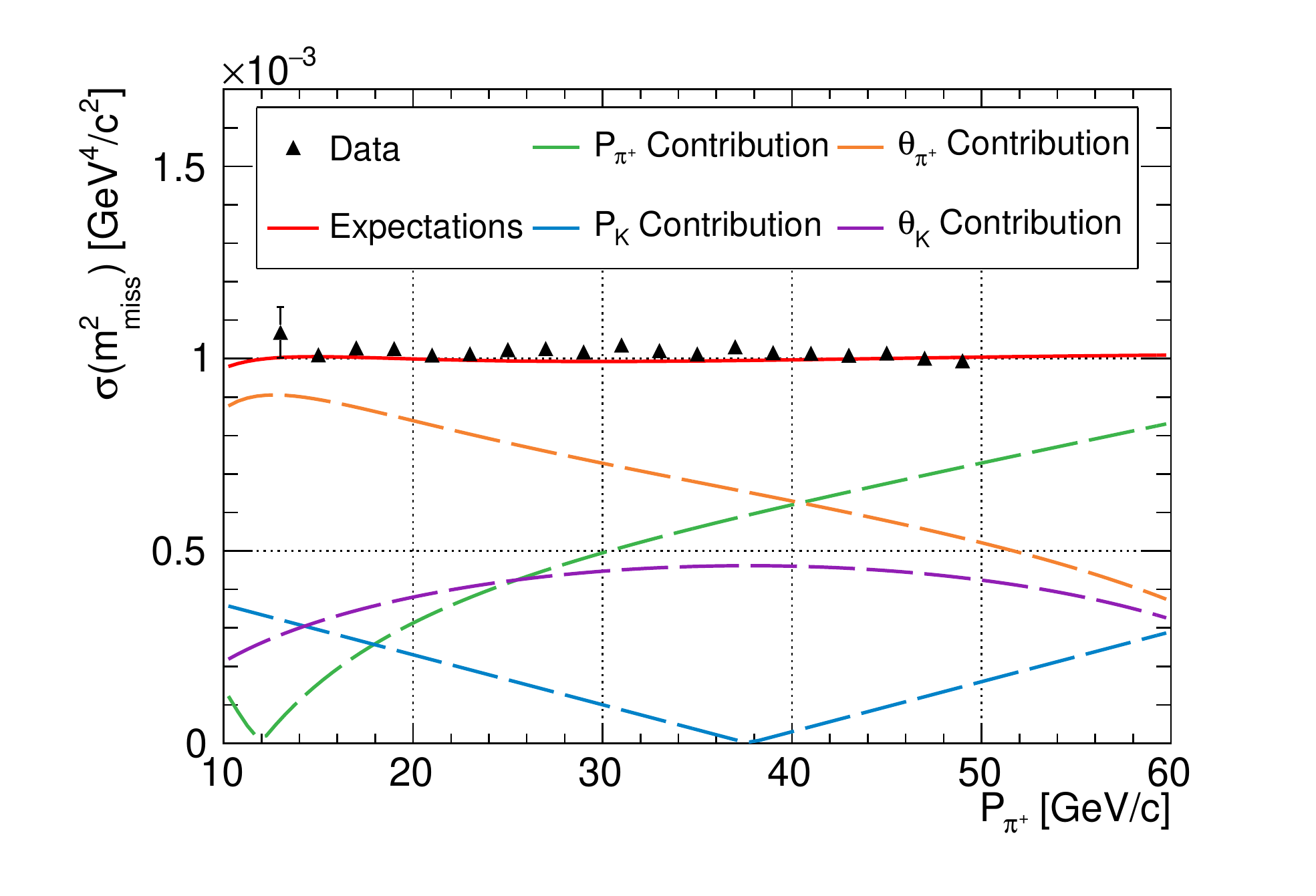}
    \caption{Squared missing mass resolution for \Kpipizero\ as function of the \Pip\ momentum for the 2016 data (black triangle) overlaid with the resolution expected based on nominal performance (plain red line) and the expected contributions (dashed line).}
    \label{fig:kine_reso}
  \end{center}\end{figure}

The figure of merit for the kinematics performance is the squared missing mass resolution for \Kpipizero\ as defined in Equation~\ref{eq:mm2}. This quantity is shown in \fig{fig:kine_reso} as a function of the \Pip\ momentum for the 2016 data overlaid with the expectations based on nominal performance. The contribution of the \Pip\ and \Kp\ momentum and angular resolutions are also shown. The overall agreement between the squared missing mass resolution obtained with a simulation based on nominal performance and the measured ones indicates that the GTK momentum and angular resolutions are matching the specifications.

\subsection{Efficiency}

The \Kpipizero\ background rejection obtained using the decay kinematics becomes ineffective when a wrong track is assigned to the \Kp. Such situations occur when \Kp\ hits are missing due to the limited GTK efficiency. The intrinsic GTK inefficiency is thus defined as the fraction of events in which hits from the \Kp\ triggering the detector readout are missing. Events for which \Kp\ hits are expected to be missing due to an identified readout inefficiency are considered as dead-time and not counted as inefficiency. Likewise, missing hits due to geometrical consideration are counted as geometrical acceptance loss. The cumulated loss due to dead-time and geometrical acceptance is about 5\%.

The efficiency was measured using samples of \Kpipipi\ selected and reconstructed in the same fashion as for the time resolution measurements. The efficiency is then derived from the fraction of reconstructed \Kp\ matched with a GTK track. A good match is realised when the GTK track and the \Kp\ have compatible momentum, direction, time and position at GTK3.

Using this method, the intrinsic GTK efficiency was found to be to be 97.0\% which, assuming no correlation and similar behaviour of the three stations, corresponds to a 99.0\% station efficiency. This performance matches the detector specifications.

\begin{figure}[!b]\begin{center}
    \resizebox{\textwidth}{!}{\begin{tikzpicture}[level 1/.style={sibling distance=55mm},level 2/.style={sibling distance=45mm},
    every node/.style = {shape=rectangle, rounded corners,
      draw, align=center,anchor=north}]
  \node (A) at (0,0) {\textcolor{blue}{Total}\\ 1687617} 
  child { node {Missing GTK Hits\\ 51191 (\textcolor{blue}{3.03}) }
    child { node {GTK1 Hit Missing\\ 16572 (\textcolor{blue}{0.98}) } 
      child { node[align=right] { Dead Pixel: 12181 (\textcolor{blue}{0.72})\\ Calibration: 0 (\textcolor{blue}{0.00})\\ Hit Collision: 3214 (\textcolor{blue}{0.19})\\ Other: 1177 (\textcolor{blue}{0.07})} }
    }
    child { node {GTK2 Hit Missing\\ 11590 (\textcolor{blue}{0.69}) } 
      child { node[align=right] { Dead Pixel: 3738 (\textcolor{blue}{0.22})\\ Calibration: 0 (\textcolor{blue}{0.00})\\ Hit Collision: 4393 (\textcolor{blue}{0.26})\\ RO Saturation: 3459 (\textcolor{blue}{0.20})} }
    }
    child { node {GTK3 Hit Missing\\ 16342 (\textcolor{blue}{0.97}) } 
      child { node[align=right] { Dead Pixel: 2137 (\textcolor{blue}{0.13})\\ Calibration: 7801 (\textcolor{blue}{0.46})\\ Hit Collision: 5483 (\textcolor{blue}{0.32})\\ Other: 921 (\textcolor{blue}{0.05})} }
    }
    child { node {$\geq$2 GTK Hit Missing\\ 6687 (\textcolor{blue}{0.40})  }}
  }
  child { node {GTK Match\\ 1636426 (\textcolor{blue}{96.97})} };
\end{tikzpicture}}
    \caption{Partition of the \Kpipipi\ collected during an eight hours data taking period in October 2017. The numbers denote the numbers of events in each category. The numbers in parenthesis are the percentage of the total number of \Kpipipi.}
    \label{fig:treeEfficiency}
  \end{center}\end{figure}
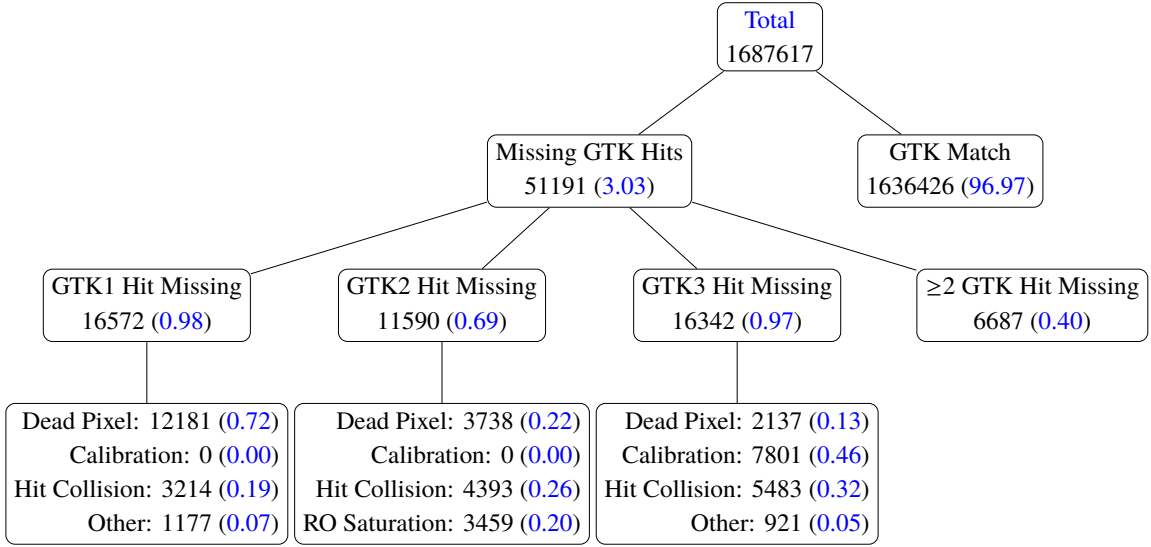

Studies were pursued to identify the origin of the missing hits. Unmatched events were reprocessed considering GTK tracks formed by hits in two stations only. Sources were identified based on the expected position of the missing station hit and searching for hits offset in space or time. The main inefficiency sources found are dead pixels, local mis-calibration inducing signal loss on pixels set at higher thresholds, and synchronous firing of pixels multiplexed in the the same time-stamping unit inducing ambiguities in the hit pixel address. Last, few hits were lost in GTK2 due to the misalignment between the beam peak intensity region and the station centre as shown in \fig{fig:hitmap}. This misalignment induces a much higher rate on few chips. At the 2016 beam intensity, this effect was irrelevant but at higher beam intensities, the readout electronics of the most illuminated chips started to saturate.
The importance of each source varies depending on the station and is reported in \fig{fig:treeEfficiency}. The stations inefficiency are slightly correlated as the position of some dead pixel clusters are aligned between two stations.

  \section{Conclusions}
  The NA62 physics case requires the development of a beam spectrometer, the GigaTracKer, with unprecedented specifications in terms of particle rate, material budget and time resolution. To fulfil those requirements, a radically novel approach to tracking, based on particle time-stamping at the sub nanosecond level, was developed. A custom made chip, the TDCPix, and a carefully designed sensor allowed for this tour-de-force.

The time-stamp based approach enables the reduction of the number of tracking stations to three, the bare minimum for a magnetic spectrometer. The detector material budget is furthermore reduced by using silicon micro-channel cooling plates, a world first development in high energy physics.

The detector was successfully assembled and operated. Performance in most cases reach or exceed the specifications. The results achieved in terms of time resolution for the single hit are \vu{115}{ps} corresponding to a track time resolution of \vu{65}{ps}.

  \section*{Acknowledgements}
  The authors are grateful to the staff of the CERN laboratory and the technical staff of participating universities and laboratories for their valuable help during the construction, installation and commissioning of the  GTK detector. The present work was completed thanks to the dedication of the whole NA62 Collaboration in operating the experiment in data-taking conditions and later in providing results from the off-line data processing. The cost of the GTK and of its auxiliary systems were supported  by the funding agencies of the collaborating Institutes. We are particularly indebted to: F.R.S.-FNRS (Fonds de la Recherche Scientifique - FNRS), Belgium; INFN (Istituto Nazionale di Fisica Nucleare), Italy; CERN (European Organization for Nuclear Research), Switzerland.
  
   \bibliographystyle{unsrt-mathi.bst}
   \bibliography{biblio.bib}

\end{document}